\documentclass[a4paper,twocolumn,english,aps,letter]{revtex4}
\usepackage[OT1]{fontenc}
\usepackage[latin1]{inputenc}
\usepackage{babel}
\usepackage{graphics}

\makeatletter

\providecommand{\LyX}{L\kern-.1667em\lower.25em\hbox{Y}\kern-.125emX\@}

\usepackage{pslatex}

\makeatother
\begin{document}

\title{Excitation gaps in fractional quantum Hall states: An exact diagonalization
study}

\author{R.H. Morf\emph{\( ^{(1)} \),} N. d'Ambrumenil\( ^{(2)} \) and S.
Das Sarma\( ^{(3)} \)}

\affiliation{\( ^{(1)} \)Condensed Matter Theory, Paul Scherrer Institute, CH-5232
Villigen, Switzerland\\
 \( ^{(2)} \)Department of Physics, The University of Warwick, Coventry,
CV4 7AL, UK\\
 \( ^{(3)} \)Department of Physics, University of Maryland, College
Park, MD 20742, USA}

\date{22 February 2002}

\begin{abstract}
We compute energy gaps for spin-polarized fractional quantum Hall
states in the lowest Landau level at filling fractions \( \nu =\frac{1}{3} \),
\( \frac{2}{5} \), \( \frac{3}{7} \) and \( \frac{4}{9} \) using
exact diagonalization of systems with up to 16 particles and extrapolation
to the infinite system-size limit. The gaps calculated for a pure
Coulomb interaction and ignoring finite width effects, disorder and
LL mixing agree well with the predictions of composite fermion theory
provided the logarithmic corrections to the effective mass are included.
This is in contrast with previous estimates, which, as we show, overestimated
the gaps at \( \nu = \)2/5 and 3/7 by around 15\%. We also study
the reduction of the gaps as a result of the non-zero width of the
2D layer. We show that these effects are accurately accounted for
using either Gaussian or `z\( \times  \) Gaussian' (zG) trial wavefunctions,
which we show are significantly better variational wavefunctions than
the Fang-Howard wavefunction. The Gaussian and zG wavefunctions give
Haldane pseudopotential parameters which agree with those of self-consistent
LDA calculations to better than \( \pm  \) 0.2\%. For quantum well
parameters typical of heterostructure samples, we find gap reductions
of around 20\%. The experimental gaps, after accounting heuristically
for disorder, are still around 40\% smaller than the computed gaps.
However, for the case of tetracene layers in metal-insulator-semiconductor
(MIS) devices we find that the measured activation gaps are close
to those we compute. We discuss possible reasons why the difference
between computed and measured activation gaps is larger in GaAs heterostructures
than MIS devices. Finally, we present new calculations using systems
with up to 18 electrons of the gap at \( \nu =\frac{5}{2} \) including
width corrections.
\end{abstract}


\maketitle

\section{Introduction}

Our understanding of the FQHE \cite{tsui82} is primarily based on
the Laughlin wavefunction (wf) \cite{laughlin83} and its appropriate
hierarchical generalizations \cite{haldane83,halperin84,jain89} for
the so-called higher order {}``daughter'' fractions which are many-electron
wavefunctions in the lowest Landau level with no adjustable parameters.
The fundamental property underlying the FQHE phenomenon is the existence,
at certain filling fractions of the lowest Landau level, of an incompressible
ground state and an energy gap \( \Delta  \) in the many-body excitation
spectrum. This gap is produced entirely by the electron-electron interaction
while the corresponding non-interacting single particle energy levels
are all degenerate at the particular fractional filling (i.e. all
non-interacting single particle levels have energy \( \hbar \omega _{c}/2 \)
in the lowest Landau level, where \( \omega _{c}=eB/(mc) \) is the
cyclotron frequency in the magnetic field \( B \)). 

The excitation gap \( \Delta  \) is the key measure of the robustness
of the FQHE - the incompressibility cannot be destroyed by weak disorder
in the system if the gap is large. The behavior of the gap as a function
of filling fraction in the main sequence of FQHE states can also be
compared to predictions of the composite fermion (CF) picture and
used to extract the CF effective mass. The excitation gap at \( \nu =1/3 \)
has been theoretically estimated on the basis of exact diagonalization
studies \cite{haldane_reza85,fano86} and Monte Carlo calculations
\cite{morf_ha86,morf_ha87} as have the gaps at filling fractions
\( \nu =2/5 \) and \( \nu =3/7 \) \cite{nda_morf89}. The numerically
computed estimates of the gap are, however, significantly larger (by
a factor of 2 to 3 at \( \nu =1/3 \) for example) than the measured
gaps, \( \Delta _{a} \), deduced from the activated temperature-dependence
of the longitudinal resistivity minimum for each fraction \cite{boebinger85,willett88,du93,du94}. 

Here we report the results of extensive finite-size studies of the
gap for spin-polarized excitations of electrons confined to the lowest
Landau level (LLL) at filling fractions \( \nu =\frac{1}{3} \), \( \frac{2}{5} \),
\( \frac{3}{7} \) and \( \frac{4}{9} \) as well as detailed results
for the quantum Hall state at \( \nu =\frac{5}{2} \). We give a detailed
analysis of the finite size corrections and show that previous estimates
of the gap \cite{nda_morf89} for the pure Coulomb interaction at
\( \nu =2/5 \) and \( 3/7 \) were around 20\% too high as a result
of inaccurate extrapolation methods. For the simple exactly solvable
case of a single hole in a filled polarized lowest Landau level, we
demonstrate how an optimized extrapolation scheme dramatically reduces
errors in the estimate for the infinite system result. We compare
our new, more accurate, results with the predictions of composite
fermion theory \cite{HLR,stern_halperin95,Bert_in_Perspectives}.
We find that, whereas previous estimates were consistent with a CF
effective mass which was independent of filling fraction, the new
estimates are in better agreement with the CF theory which predicts
a logarithmically divergent effective mass as a function of filling
factor as \( \nu =1/2 \) is approached \cite{HLR,stern_halperin95}.
The new results are also closer to the estimates of the effective
mass from another type of finite-size calculations at \( \nu =1/2 \)
\cite{morf_da95}.

Other previous larger estimates of the pure Coulomb gap \cite{jain_kamilla97}
may also involve an inaccurate extrapolation to the infinite system
limit, but as these results were obtained using CF trial wf's and
Monte Carlo techniques we cannot say for certain where the origin
of this difference lies. However, we mention that our calculated excitation
gaps are lower by as much as 30\% than those in \cite{jain_kamilla97},
and some discrepancy exists even for the pure Coulomb interaction
results at \( \nu =1/3 \) where our extrapolation to the thermodynamic
limit is most reliable. 

We clarify to what extent the discrepancy between numerically computed
gaps and those extracted from transport measurements can be attributed
to finite-width effects. The large disagreement between experimental
activation gaps \( \Delta _{a} \) and the numerically computed gaps
\( \Delta _{c} \) has been an outstanding problem in the subject
since the first accurate measurement of activation gaps was reported
more than fifteen years ago \cite{boebinger85}. There have been several
previous theoretical attempts to compute realistic estimates of the
energy gap and to identify the source of the large discrepancy between
\( \Delta _{c} \) and \( \Delta _{a} \) \cite{zhang_das86,yoshioka86,ortalano97,park_j99,morf_park_j99,park_mj99}.
These took account of the finite thickness correction (i.e. relaxing
the pure \( 1/r \) Coulomb interaction approximation by including
the softening introduced by the transverse width of the 2D layer),
and of the Landau level mixing corrections \cite{yoshioka84,zhu_louie93}.
There have also been studies of the spin-reversed excitations which
are the lowest lying excitations for small g-factors and small magnetic
fields \cite{halperin_hpa,chakraborty_86,morf_ha87,rezayi_spin}. 

There are reports in the literature \cite{ortalano97} that the finite-width
effects account for all the difference between measured and theoretically
predicted gaps. Our results are at variance with this conclusion \cite{ortalano97}
and consequently also with the results of \cite{park_j99} which were
based on incorrect results from \cite{ortalano97}. The error in \cite{park_j99}
was originally corrected in \cite{morf_park_j99,park_mj99}. We find
on the basis of the largest finite-size diagonalizations to date and
of a careful analysis of the finite-size corrections that the finite-width
corrections account for at most half of the difference between the
computed gaps and those observed in GaAs heterostructures. On the
other hand the gaps observed recently in tetracene in metal insulator
semiconductor structures are only slightly smaller than our estimates.
We discuss the possible reasons for these discrepancies. We argue
that they are unlikely to be due to spin-reversed excitations or Landau-level
mixing and suggest that they are the result of disorder effects which
may affect the activation energy for transport differently in heterostructures
and MIS devices.

We show that it is unlikely for there to be a transition from an incompressible
to a compressible state at fixed filling factor, for 
\( \nu =\frac{1}{3},\frac{2}{5},\frac{3}{7} \),
caused by a gap collapse induced entirely by the softening of the
Coulomb interaction due to the finite thickness corrections. Such
a transition has been conjectured to occur in the second Landau level
\cite{morf98,Rez-Hald00}, where the FQHE is much less robust. It
may also happen in situations where increasing the width in the transverse
direction changes the symmetry of the sub-band wf \cite{princeton}.
Alternatively, a new kind of FQH state can arise in square, parabolic
or double wells, where, for large enough well width, the wf may split
into an effective double layer structure at the two ends of the well
with a central self-consistent barrier separating these two effective
layers \cite{princeton-bell}. In the regular GaAs heterostructure
system \cite{willett88,du93,du94,boebinger85} we find the lowest
Landau level \( \nu =\frac{1}{3},\frac{2}{5},\frac{3}{7} \) FQHE
to be robust with respect to the finite thickness effect, with \( \Delta >0 \)
even for the largest possible (and physically allowed) transverse
thickness. However the actual value of \( \Delta  \) may become rather
small and one might have to go to very low temperatures (and very
high quality, low disorder samples) to observe the FQHE. Our results
are in conflict with the claim by Park et al. \cite{park_j99,park_mj99,songhe90}
that the finite width alone can lead to the loss of incompressibility
at a filling fraction \( \nu =p/(2p+1) \) for some finite value \( p=p_{c} \).

We also compare the various approximate methods for accounting for
finite thickness effects based on interfacial trial wf's with those
taken from self-consistent local density approximation (SCLDA) calculations
\cite{ortalano97,stern_das84}. Previous model calculations have used
Gaussian and Fang-Howard envelope wf's and the Zhang-DasSarma (ZDS)
model interaction \cite{zhang_das86}. We introduce a new variational
envelope wf, the {}``z \( \times  \)Gaussian'' (zG). We find that
both the zG and the Gaussian envelope wf's give Haldane pseudopotential
parameters which agree to within fractions of a percent with those
from the full SCLDA wf's with the zG giving slightly more accurate
results at the densities used in experiment. However, both give essentially
indistinguishable results for excitation energies and gaps from those
taken from the SCLDA wf's. This result shows that accurate finite-size
studies of finite-width effects require only the determination of
the appropriate width parameter in either the Gaussian or zG description
and do not require the use of SCLDA based tables of parameters \cite{ortalano97}. 

We show that depending on the sub-band density either the Gaussian
or zG variational wf provide substantial quantitative improvements
over the well-known Fang-Howard variational wf \cite{ando_fs82} which
has been employed extensively in heterostructure electronic calculations.
Indeed, it turns out that the Fang-Howard wf generally overestimates
the kinetic energy, and consequently predicts significantly too large
width. The expectation value of the energy and other quantities of
interest in this context can be calculated analytically for these
variational wf's, and in the case of the Gaussian, it is easy to perform
expansions for either very small or very large width. Finally in the
Appendix, we explain why the ZDS model is not reliable directly for
predicting finite thickness corrections, but we present a simple modification
which corrects its main shortcoming.

The remainder of this paper is organized as follows: In section II
we describe the diagonalization of the \( N \)-particle Hamiltonian
in the spherical geometry and give the definitions of the quasiparticle,
quasihole and gap energies. In section III, we discuss the extrapolation
to the \( N\rightarrow \infty  \) limit and in section IV we compare
the variation of the calculated gaps with filling fraction \( \nu  \)
with the predictions of composite fermion theory \cite{HLR,stern_halperin95,Bert_in_Perspectives}.
In section V we show how variational wf's can be used to model finite
width effects. In section VI, we compute the reduction of the energy
gaps as a function of the finite width and in section VII we compare
our results for the gap energies with experimentally reported estimates
of gaps.

\section{Quasiparticle and Quasihole Energies}

We model the two-dimensional electron gas using Haldane's spherical
geometry \cite{haldane83}. Particles with coordinates \( (R,\theta _{i},\phi _{i}) \)
move in a monopolar magnetic field of strength \( B=S\, \hbar /eR^{2} \)
which gives rise to \( 2S+1 \) linearly independent cyclotron orbits
in the lowest Landau level. The single particle orbitals on the surface
of the sphere for the particles in the lowest Landau level are then
functions \( \psi (\theta _{i},\phi _{i}) \) which are the lowest
energy eigen states of the kinetic energy.

In the lowest Landau level, the interaction between particles is written
\begin{equation}
\label{Pseudopots}
V(ij)=\sum _{m}\sum ^{N}_{i<j}V_{m}P_{m}(ij)
\end{equation}
 where \( P_{m}(ij) \) projects onto states in which particles \( i \)
and \( j \) have relative angular momentum \( m\hbar  \) and \( V_{m} \)
gives their interaction energy for this relative angular momentum.
The set \( V_{m} \), called Haldane pseudopotentials \cite{haldane83},
completely characterizes the interaction between particles confined
to the lowest Landau level. In terms of the electron-electron interaction,
\( V(r) \), they are defined in the plane by \cite{Haldane_in_QHE}

\begin{equation}
\label{eq:pseudopotential}
V_{m}^{(n)}=\frac{1}{(2\pi )^{2}}\int d\vec{r}V(r)\int d\vec{q}\, e^{i\vec{q}.\vec{r}-(q\ell _{0})^{2}}\, \Big (L_{n}(\frac{q^{2}\ell _{0}^{2}}{2})\Big )^{2}\, L_{m}(q^{2}\ell _{0}^{2}),
\end{equation}
where \( n \) refers to the Landau level and \( V(r) \) stands for
the electron electron interaction. The corresponding integrals for
electrons on the surface of a sphere are described in \cite{haldane_reza85}.
In the lowest Landau level \( n=0 \), the first Laguerre polynomial
in equation (\ref{eq:pseudopotential}) is equal to unity. As we shall
discuss in the next section, the effect of the finite width of the
wf, \( \phi (R_{i}), \) is incorporated in these pseudopotential
parameters \( V^{(n)}_{m} \). In the following, we will drop the
superscript \( n=0 \) and denote the Haldane pseudopotentials for
the lowest Landau level by \( V_{m} \).

The method for computing excitation energies and gaps in this geometry
has been described in detail in many places \cite{haldane83,fano86,nda_morf89}.
According to the hierarchy model, the FQHE ground states at filling
fraction \( \nu  \) occur for a system of \( N \) particles when
the total flux \( 2S \) is given by

\begin{equation}
\label{2S(nu,N)}
2S_{0}(\nu ,N)=\nu ^{-1}N+X(\nu )
\end{equation}
 where \( X(\nu ) \) is the shift function \cite{nda_morf89}, which
is a characteristic of the geometry of the system (in this case the
sphere) \cite{wenshift}. Laughlin's \cite{laughlin83} elementary
fractionally charged excitations from the FQHE ground state at filling
fraction \( \nu =p/(2p+1) \) correspond to the ground state configuration
of a system with additional/missing flux \( \pm 1/p \), \begin{equation}
\label{2Sqpqh}
2S_{\pm 1/p}(\nu ,N)=2S_{0}(\nu ,N)\pm \frac{1}{p}.
\end{equation}
 At \( \nu =1/m \) there are systems with both \( 2S_{0} \) and
\( 2S_{\pm 1/p} \) both integer for all integer \( N. \) At other
filling fractions \( 2S_{0} \) and \( 2S_{\pm 1/p} \) are never
both integer for the same \( N \). For example at \( \nu =3/7, \)
\( 2S_{0} \) is integer when the particle number is \( N=3n \) (\( n \)
integer) while \( 2S_{\pm 1/3} \) is integer for \( N=3n\mp 1 \),
respectively. We take the energy to nucleate a single quasiparticle/quasihole
in a system of N particles at filling fraction \( \nu  \), \( e_{\nu }^{\pm }(N) \),
to be the total energy difference between the lowest energy state
with total flux \( 2S_{\pm 1/p}(\nu ,N) \) and the total ground state
energy the system would have at \( 2S_{0}(\nu ,N) \) for the same
\( N \), i.e. \begin{equation}
\label{proper energy}
e_{\nu }^{\pm }(N)=E_{2S_{\pm 1/p}}(N)-E_{2S_{0}}(N).
\end{equation}
Here \( E_{2S\pm 1/p}(N) \) is the total energy of the system of
\( N \) particles in their ground state in \( 2S_{\pm 1/p} \) flux
quanta. For filling fractions \( \nu =1/m \) we can calculate both
energies directly, while at filling fractions \( \nu =p/(2p+1) \)
with \( p\neq 1 \) we have to estimate \( E_{2S_{0}}(N) \) by interpolating
(or extrapolating for the largest system sizes) between system sizes
for which we can compute \( E_{2S_{0}}(N) \).

\begin{figure}
{\centering \resizebox*{8cm}{!}{\includegraphics{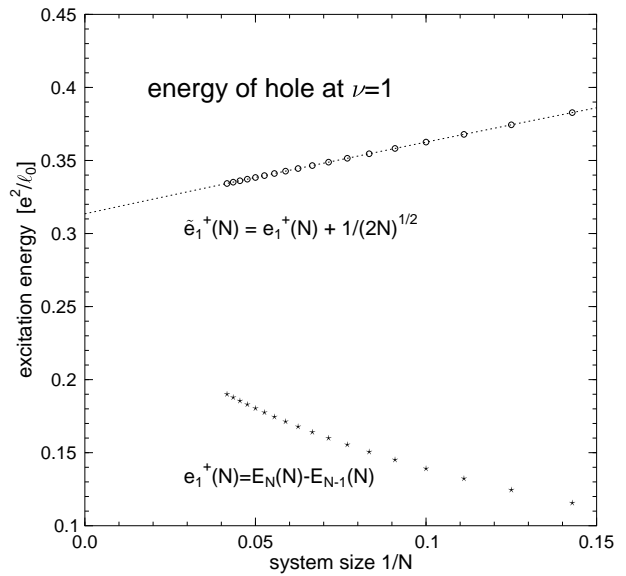}} \par}

\caption{\label{hole at nu=3D1}The energy of a single hole. The asterisks
show \protect\( e_{1}^{+}(N)\protect \) and the circles show the
corrected energy \protect\( \widetilde{e}_{1}^{+}(N)\protect \).
The extrapolation in \protect\( 1/N\protect \) of the corrected energies
pass through the exact value \protect\( \frac{1}{4}\sqrt{\pi /2}\protect \)
whereas extrapolation of \protect\( \epsilon _{1}^{+}(N)\protect \)
in \protect\( 1/N\protect \) would give incorrect results.}
\end{figure}

\section{Energy Gaps: Finite Size Effects and the Thermodynamic Limit}

From studying the variation with system size of the
energies to nucleate quasiparticles and quasiholes in finite-size
systems we estimate the excitation energies in the thermodynamic limit.
It is essential that the extrapolation procedure is carried out carefully.
Firstly, if working with the Coulomb interaction it is usual to quote
energies in units of \( e^{2}/\varepsilon \ell '_{_{0}} \), where
\( \ell '_{0}=\sqrt{\hbar c/eB} \) is the magnetic length and \( \varepsilon  \)
is the dielectric constant for the medium. However, for systems with
number density \( n_{S} \) on the sphere, \( \ell '_{0}=\sqrt{(1/(2\pi n_{S}))(N/(2S))} \)
and so for systems at fixed density, the magnetic length \( \ell '_{0} \)
depends on the particle number and the total flux through the ratio
\( N/2S \). In order to compare quantities measured in the same units
we convert all energies by using the magnetic length in the infinite
system \( \ell _{0} \) = \( \sqrt{\nu /(2\pi n_{S})} \) . 

There is also a systematic contribution to the excitation energy in
a finite size system which scales to zero in the thermodynamic limit,
which we can take account of explicitly \cite{nda_morf89}. When the
localized quasiparticle/quasihole excitation which is formed in a
system of \( N \) particles around the point on the sphere \( R\vec{\Omega } \)
with \( \vec{\Omega } \) a unit vector pointing away from the origin
a charge \( \pm qe \) with \( q=1/(2p+1) \) is concentrated around
this point. This charge has come from the rest of the system. There
is then a contribution, \( A_{q} \), to the energy of the system
from the non-uniform distribution of charge on the surface of the
sphere which, in units of \( e^{2}/\varepsilon \ell _{0} \), is given
by\begin{equation}
\label{eq:aq}
A_{q}(\nu )=-q^{2}\sqrt{\frac{\nu }{2N}}.
\end{equation}
To extrapolate to the infinite system size limit it is better to remove
this contribution explicitly and study the corrected quasihole and
quasiparticle energies

\begin{equation}
\label{epstilde}
\widetilde{e}_{\nu }^{\pm }(N)\equiv e_{\nu }^{\pm }(N)-A_{q}(\nu ).
\end{equation}
We also define the corrected gap energies to be the sum of quasiparticle
and quasihole energies\begin{equation}
\label{gap(N)}
\widetilde{e}_{\nu }^{g}(N)\equiv \widetilde{e}_{\nu }^{+}(N)+\widetilde{e}_{\nu }^{-}(N).
\end{equation}
We denote the limit \( N\rightarrow \infty  \) of the gap 
and
quasihole, quasiparticle excitation energies by 
\( \widetilde{\epsilon }_{\nu }^{(g)} \) and
\( \widetilde{\epsilon }_{\nu }^{\pm } \) respectively.

To illustrate the importance of working with these corrected energies,
we show results for a single hole at \( \nu =1 \), which is a case
for which the energy can be computed analytically using the exact
expression for the energy of a filled Landau level \cite{fano88}.
We find\begin{equation}
\label{hole_at_nu=1}
e_{1}^{+}(N)=-\frac{1}{2}\frac{E_{N-1}(N)}{N}\left( 1+\frac{3}{2N+1}\right) -\frac{1}{\sqrt{2N}}.
\end{equation}
 The contribution \( -1/\sqrt{2N} \) is just the correction \( A_{1}(1) \)
for the case \( \nu =1. \) Both \( e_{1}^{+}(N) \) and \( \widetilde{e}_{1}^{+}(N) \)
are shown as a function \( 1/N \) in Figure \ref{hole at nu=3D1}.
It is clear from the figure that extrapolation of \( e_{1}^{+}(N) \)
with \( 1/N \) would give spurious results because of the contribution
of \( A_{1}(1) \), which varies as \( 1/\sqrt{N} \). By contrast,
extrapolation of \( \widetilde{e}_{1}^{+}(N) \) with \( 1/N \) gives
the correct result \cite{jancovici} \( \lim _{N\rightarrow \infty }(-E_{N-1}(N)/(2N))=\frac{1}{4}\sqrt{\pi /2} \).
\vspace{0.3cm}

\begin{figure}
{\centering \resizebox*{8.4cm}{!}{\includegraphics{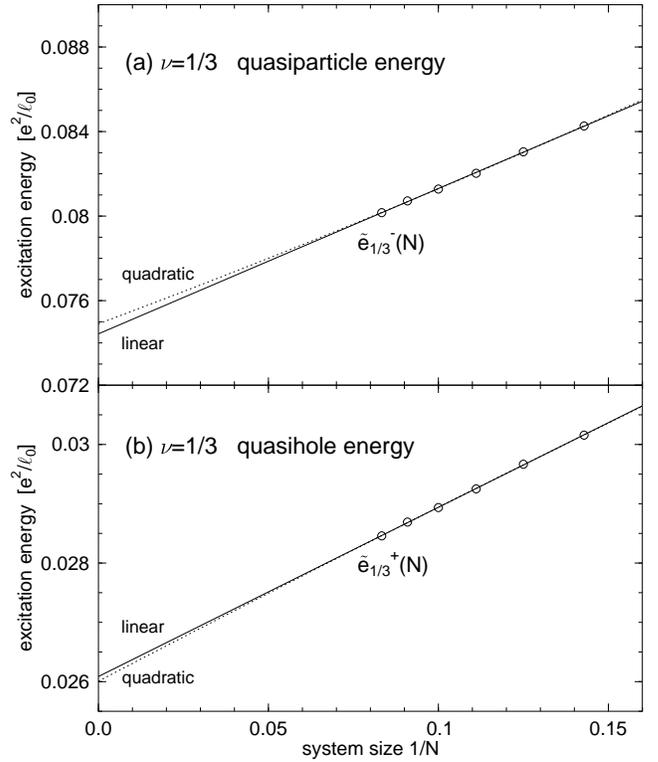}} \par}

\caption{\label{nu3_qpqh} The quasiparticle (\protect\( \widetilde{e}_{1/3}^{-}(N)\protect \))
and quasihole (\protect\( \widetilde{e}_{1/3}^{+}(N)\protect \))
energies at \protect\( \nu =1/3\protect \) and the extrapolation
using linear and quadratic functions of \protect\( 1/N\protect \).
We take the small differences in the extrapolated result as a measure
of the accuracy of the extrapolation. }
\end{figure}

\begin{figure}
{\centering \resizebox*{8cm}{!}{\includegraphics{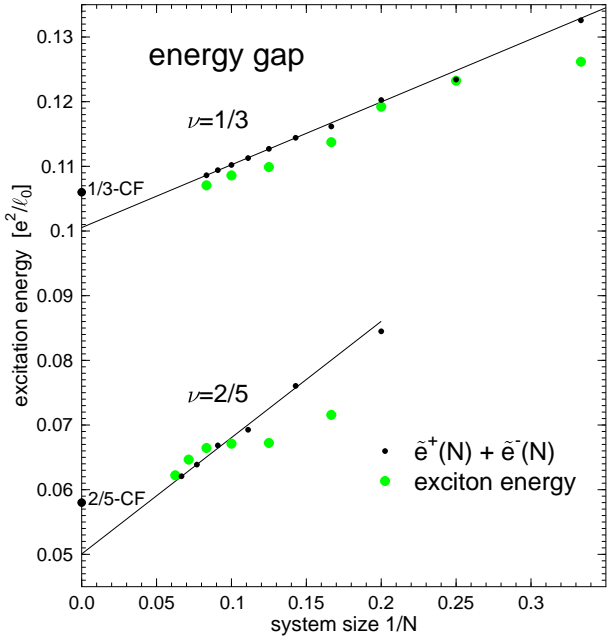}} \par}

\caption{\label{gaps1/3and2/5}The gap energies \protect\( \widetilde{e}^{g}_{\nu }(N)\protect \)
at \protect\( \nu =1/3\protect \) and \protect\( 2/5\protect \),
computed using (\ref{gap(N)}) (small solid dots), and the neutral
exciton energies for \protect\( L=N\protect \) at \protect\( \nu =1/3\protect \)
and \protect\( L=(N+2)/2\protect \) at \protect\( \nu =2/5\protect \)
(large shaded dots)\protect\( .\protect \) Also shown by circles
on the vertical axes are the estimates of the gap energies of Jain
and Kamilla \cite{jain_kamilla97} obtained from an analysis of trial
composite fermion wf's. The straight lines denote the best linear
(in \protect\( 1/N\protect \)) fit to the data points. The intercepts
give the estimate of the gap energy \protect\( \widetilde{e}^{g}_{\nu }(\infty )\protect \)
neglecting corrections due to non-zero width effects and higher Landau
levels.}
\end{figure}

In Figure \ref{nu3_qpqh} we show \( \widetilde{e}_{1/3}^{+}(N) \)
and \( \widetilde{e}_{1/3}^{-}(N) \) and their extrapolations to
\( N=\infty  \) using least squares fits to linear and quadratic
functions in \( 1/N \). We take the difference, 0.0005 for the quasiparticle
and 0.0002 for the quasihole, in the estimates from the two different
extrapolation procedures as our measures of the accuracy of the extrapolation.
In fact, inclusion of the \( 1/N^{2} \) term in the fit does not
improve the \( \chi ^{2} \) value significantly. So, in the following,
we will always use linear extrapolation in \( 1/N \) to compute gaps
in the thermodynamic limit. Figure \ref{gaps1/3and2/5} shows the
gap energies at \( \nu =1/3 \) and \( 2/5 \) as functions of \( 1/N \)
and the extrapolations to \( N=\infty  \) limit together with the
estimates based on the study of trial CF-wf's \cite{jain_kamilla97}.
Plotted are the sum of quasiparticle and quasihole energies \( \tilde{e}^{-}(N)+\tilde{e}^{+}(N) \),
using the correction \( A_{q}(\nu ) \) (\ref{eq:aq}) and the energy
\( \tilde{e}_{exc}(\nu ) \) of the neutral excitation with \( L=L_{max} \)
(cf. caption to Figure \ref{gaps1/3and2/5}), corresponding to maximum
separation of the quasiparticle and quasihole on the sphere, again
corrected by the term \( A_{q}(\nu ) \) which stands for the Coulomb
energy between the quasiparticle and quasihole. As can be seen, the
size dependence of the exciton energies is much less smooth than that
of the sum of qp- and qh-energies. Indeed, if they were known only
for small systems, extrapolation to the bulk limit would be inaccurate.
Only for the largest systems, does the size dependence of exciton energies
become smooth and allow reliable extrapolation to the thermodynamic
limit, which is consistent with that based on the sum of qp- and qh-energies,
although less accurate (see Figure \ref{gaps1/3and2/5}).

\begin{figure}
{\centering \resizebox*{7.9cm}{!}{\includegraphics{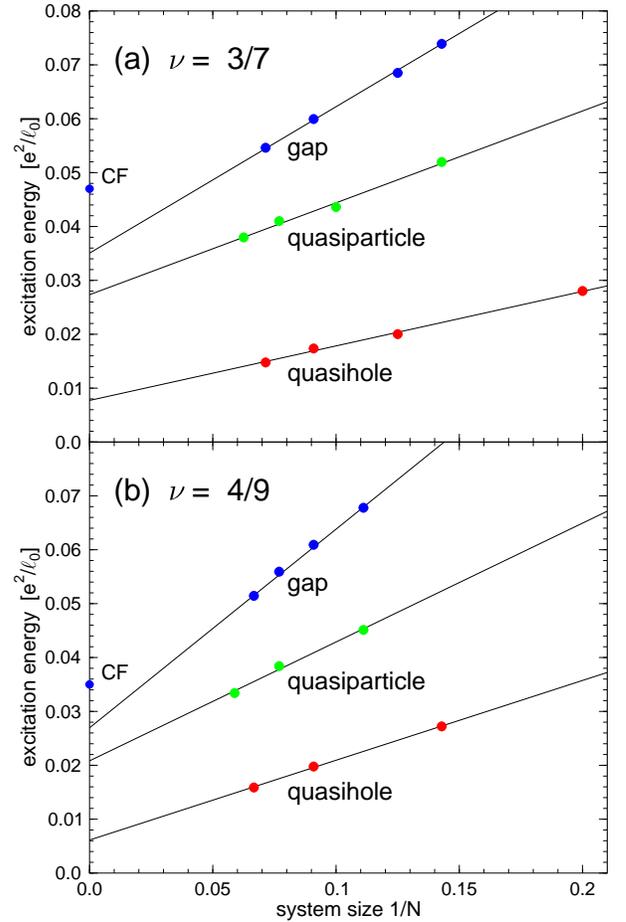}} \par}

\caption{\label{gaps3/7and4/9} The quasiparticle and quasihole energies \protect\( \widetilde{e}^{\pm }_{\nu }(N)\protect \)
at \protect\( \nu =3/7\protect \) and \protect\( 4/9\protect \)
and the best linear fits to these points. The sum of the two linear
functions can be taken as a measure of the gap energies \protect\( \widetilde{e}^{g}_{\nu }(N)\protect \)
(these cannot be computed directly for these filling fractions as
the systems with single quasiparticle and quasiholes have different
numbers of particles). }
\end{figure}

Figure \ref{gaps3/7and4/9} shows the quasiparticle and quasihole
energies at \( \nu =3/7 \) and \( 4/9 \). Although the estimate
at \( \nu =1/3 \) is close to the values quoted previously \cite{nda_morf89},
the values at \( \nu =2/5 \) and \( 3/7 \) are around 20\% smaller
although still within the large uncertainties of the earlier calculation.
Our latest estimates are more accurate as a consequence of a better
understanding of finite size effects in addition to being able to
diagonalize the Hamiltonians for larger systems (with up to around
100 million basis states) - 15 particles instead of 11 particles for
the quasiparticle and quasihole at \( \nu =2/5 \) and 16 particles
instead of 13 for the quasiparticle at \( \nu =3/7 \). It is interesting
to note that the extrapolation of quasiparticle and quasihole energies
at \( \nu =1/3 \) based on small sizes (\( N=4,5,6 \)) yields the
values \( \tilde{\epsilon }_{1/3}^{-}\approx 0.0757 \), \( \tilde{\epsilon }_{1/3}^{+}\approx 0.0267 \)
and for the gap \( \Delta _{1/3}\approx 0.1024 \). These are within
about one percent of our best estimates of the bulk limit of \( \tilde{\epsilon }_{1/3}^{-}=0.0749 \),
\( \tilde{\epsilon }_{1/3}^{+}=0.0263 \) and for the gap \( \Delta _{1/3}=0.1012 \),
obtained using system sizes up to \( N=12 \) and performing the extrapolation
by linear polynomial fit in \( 1/N \). Likewise at \( \nu =2/5 \),
extrapolation using the results at \( N=5,7,9 \) yields values for
the bulk limit for the quasiparticle and quasihole energies of \( \tilde{\epsilon }_{2/5}^{-}\approx 0.0431 \)
and \( \tilde{\epsilon }_{2/5}^{+}\approx 0.00920 \) and a value
for the gap \( \Delta _{2/5}\approx 0.0523 \), while our best estimate
based on system sizes \( 7\leq N\leq 15 \) are \( \tilde{\epsilon }_{2/5}^{-}=0.0398 \),
\( \tilde{\epsilon }_{2/5}^{+}=0.0102 \) and the gap \( \Delta _{2/5}=0.0500 \),
corresponding to a difference for the gap of about 5 percent. This
observation makes us confident that it is now also possible to compute
reliable bulk limit values for the gaps at \( \nu =3/7 \) and \( 4/9 \).
Our values are \( \Delta _{3/7}=0.035 \) and \( \Delta _{4/9}=0.027 \).
The systems at \( \nu =4/9 \) were inaccessible in our earlier work
\cite{nda_morf89}.

\section{Effective Mass of Composite Fermions}

\begin{figure}
{\centering \resizebox*{7.8cm}{!}{\includegraphics{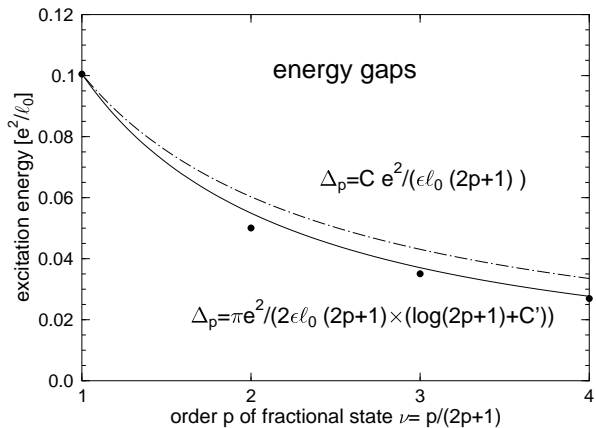}} \par}

\caption{\label{fig:eff_mass(p)}The gap energy as a function of level in
the hierarchy, \protect\( p\protect \). The estimates based on our
finite-size studies of systems with \protect\( \nu =1/3\protect \),
\protect\( 2/5\protect \), \protect\( 3/7\protect \) and \protect\( 4/9\protect \)
are shown as dots. The lower (upper) curve shows the prediction of
the CF theory with (without) and logarithmic corrections from (\ref{Gaps_in_CStheory}).
The constants \protect\( C\protect \) and \protect\( C'\protect \)
are chosen to give the correct gap at \protect\( p=1\protect \) (\protect\( \nu =1/3)\protect \).}
\end{figure}

Our estimates of the gap energies in the sequence of states \( \nu =p/(2p+1) \)
are compared in Figure \ref{eff_mass(p)} with the predictions of
CF theory \cite{HLR,stern_halperin95,Bert_in_Perspectives}, which
for this sequence gives (in units of \( e^{2}/\varepsilon \ell _{0} \))
\begin{equation}
\label{Gaps_in_CStheory}
\tilde{\epsilon }_{\nu }^{g}\equiv \widetilde{e}_{\nu }^{g}(\infty )=\frac{\pi }{2}\frac{1}{|2p+1|\left( \ln |2p+1|+C'\right) }.
\end{equation}
 Choosing \( C'=4.11 \) to fit the gap at \( \nu =1/3 \) gives the
gaps at \( \nu =2/5 \), \( 3/7 \) and \( 4/9 \) to be 0.0549, 0.0371
and 0.0276 which are remarkably close to the estimates we obtain.
We also note that, whereas the earlier estimates were better fitted
by assuming that the gaps were simply proportional to \( 1/(2p+1) \)
(i.e. ignoring the logarithmic corrections), Figure \ref{fig:eff_mass(p)}
shows that the new results are better described by the theory when
the logarithmic corrections to the gap are included. This translates
into an effective mass \begin{equation}
\label{eff_mass(p)}
m^{*}(p)=\hbar ^{2}\left( \frac{\varepsilon }{e^{2}\ell _{0}}\right) F(p)
\end{equation}
 where\begin{equation}
\label{Fofp}
F(p)=\frac{2}{\pi }\left[ \ln |2p+1|+4.11\right] .
\end{equation}

The effective mass of CF's has also been estimated by studying the
variation with system size of the ground state energy for systems
of electrons close to \( \nu =1/2 \) on the sphere with \( 2S_{0}(1/2,N)=2N-2 \)
\cite{morf_da95}. These studies gave \( F\sim 5, \) which is about
25\% larger than the value we obtain for \( p=4 \). One would expect
that, in systems close to \( \nu =1/2 \), the effective mass would
be larger than at \( \nu =4/9 \) but still finite as the long-wavelength
fluctuations of the Chern-Simons gauge field, which give rise to the
logarithmic divergence in the effective mass, will be cut off by the
level spacing.

In \cite{nda_morf89} estimates of the gap energies based on collective
excitations were also presented. On a sphere the effective wavevector
of a collective excitation is \( k_{\mbox {\it eff}}=L/R \) where
\( L \) is the total angular momentum of the system. This lowest-lying
collective excitation should correspond to a well-separated quasiparticle-quasihole
pair in the limit of large \( L. \) In the hierarchy picture, the
separation of the particle and hole should be \( 2RL/N \), so the
maximum separation possible occurs when \( L=N_{i} \). Here \( N_{i} \)
is the number of particles in the condensate of the highest (\( i \)'th)
level of the hierarchy that occurs: \( N_{0}=N \), \( N_{1}=(N+2)/2 \),
\( N_{2}=(N+6)/3 \) and \( N_{3}=(N+12)/4 \) for \( \nu =1/3 \),
\( 2/5 \), \( 3/7 \) and \( 4/9 \) respectively \cite{haldane83,haldane_reza85,nda_morf89}.
Extrapolations to the infinite system limit of the \( L=N \) excitations
should therefore give an estimate of the gap energies. The results
for \( \nu =1/3 \) and \( 2/5 \) are also included in Fig \ref{gaps1/3and2/5}.
It is clear from the figure that an extrapolation based on the exciton
energies would not be as smooth as that based on the charged excitations. 

\begin{figure}
{\centering \resizebox*{8cm}{!}{\includegraphics{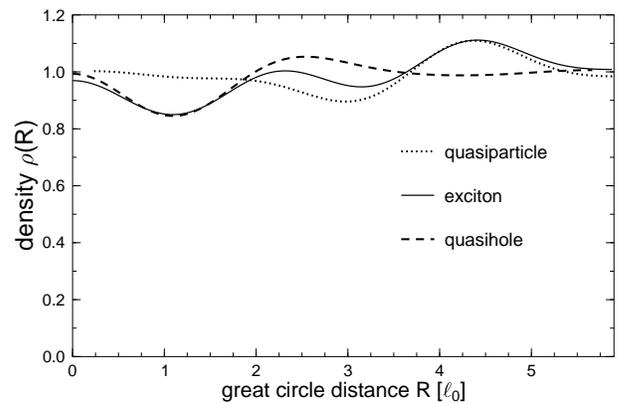}} \par}

\caption{\label{densities_qpqh_exciton}Density profiles of excitations at
\protect\( \nu =2/5\protect \) as a function of great circle distance
from the north pole: The \protect\( L=8\protect \) collective excitation
(exciton) for the 14 particle system, the quasihole (\protect\( L=7/2\protect \))
and the quasiparticle (\protect\( L=4\protect \)) for the 13 particle
system. The projection of angular momentum is maximal (\protect\( L_{z}=L\protect \))
in all cases. The origin for the quasiparticle has been shifted, so
that the point at the south pole coincides with that for the exciton.
In the thermodynamic limit, the exciton with these quantum numbers
becomes a quasihole localized at the north pole and a quasiparticle
localized at the south pole. It is clear that, even with 14 particles
at \protect\( \nu =2/5\protect \), there is still significant overlap
between the density variations associated the quasiparticle and the
quasihole localized about opposite poles. This probably explains the
large finite size effects seen (Figure \ref{gaps1/3and2/5}) in the
exciton energy as a function of \protect\( N\protect \).}
\end{figure}

We believe that the exciton energies in the small systems accessible
to direct diagonalization are not as reliable a basis for extracting
estimates of the gaps as the sum of the quasihole and quasiparticle
energies. The principal reason for this is that the quasihole and
quasiparticle states are actually ground state configurations of \( N \)
particles in total flux \( 2S_{\pm 1/p} \) and they are well-separated
in energy from all excitations. On the other hand, although the neutral
excitations are minimum energy states for the quantum numbers concerned,
they are close to the continuum of excitations for these quantum numbers
and this gives scope for large finite size effects, in addition to
leading to poor convergence and numerical instability of vector iteration
(Lanczos type) diagonalization methods. With the possible exception
of the systems at \( \nu =1/3 \), it is also clear that the system
sizes accessible to direct diagonalization are not large enough to
accommodate two excitations without significant overlap of the charge
profiles of the quasiparticle and quasihole. In Figure \ref{densities_qpqh_exciton},
we show the density profile of the 14 particle exciton at \( \nu =2/5 \),
with the corresponding quasiparticle and quasihole density profiles
for a 13 particle system overlaid for comparison. The quasiparticle
and quasihole at the opposite poles are clearly visible, but the system
is not large enough for the density profiles not to interfere.

\section{Interfacial Wavefunction and Modified Interaction}

The finite width of the sub-band envelope wf in the direction perpendicular
to the plane of the two-dimensional electron gas can be incorporated
into an effective interaction between electrons in the plane. With
the magnetic field perpendicular to the plane the single-particle
orbitals can be written:

\begin{equation}
\label{Full 3D wavefunction}
\Psi (x,y,z)=\zeta (z)\psi (x,y).
\end{equation}
The in plane wf's \( \psi (x,y) \) are eigenfunctions of the free
electron Hamiltonian in a perpendicular magnetic field, while \( \zeta (z) \)
satisfies the Schrödinger equation for a particle in the confining
potential of the quantum well or heterostructure \cite{ortalano97}.
The effective interaction between particles \( V(\vec{|}\vec{r}_{1}-\vec{r}_{2}|) \)
at positions \( \vec{r}_{1}=(x_{1},y_{1}) \) and \( \vec{r}_{2} \)
in the two-dimensional electron gas is then given by \begin{equation}
\label{2D from 3D interaction}
V(|\vec{r}_{1}-\vec{r}_{2}|)=(e^{2}/\varepsilon )\int dz_{1}\int dz_{2}\frac{\left| \zeta (z_{1})\right| ^{2}\left| \zeta (z_{2})\right| ^{2}}{\sqrt{(\vec{r}_{1}-\vec{r}_{2})^{2}+(z_{1}-z_{2})^{2}}}
\end{equation}

The study of finite-size systems is based on exact diagonalization
or the study of variational trial wf's for particles in a given Landau
level with the interparticle interaction taken to be \( V(|\vec{r}_{1}-\vec{r}_{2}|) \).
For particles on a sphere, the interaction \( V(|\vec{r}_{1}-\vec{r}_{2}|) \)
projected onto a given Landau level is characterized by Haldane's
pseudopotential parameters \( \{V_{m}\} \) (\( m=0,1,\ldots  \)).
Once these are known the exact diagonalization proceeds exactly as
in the zero-width case. (We note that as the pseudopotential parameters
are computed from the effective interaction, which assumes a planar
geometry, there is no attempt to account for any effects of the curvature
of the sphere on the finite width effects.)

Within a local density functional scheme the wf \( \zeta (z) \) satisfies
the equation\begin{equation}
\label{Schrodinger for zeta(z)}
\left( -\frac{\hbar ^{2}}{2}\frac{d}{dz}\frac{1}{m^{*}(z)}\frac{d}{dz}+V_{eff}(z)\right) \zeta (z)=E\zeta (z)
\end{equation}
 where \( V_{\mbox {eff}} \) includes the effect of the confinement
potential (including the effect of the depletion layer), the Hartree
self-interaction and exchange-correlation. For GaAs-GaAlAs quantum
wells the jump in 
\( m^{*} \) and the dielectric constant, \( \varepsilon  \)
across the interface are small and to a good approximation
both quantities can be taken to be independent of 
of \( z \) (see Table I in \cite{stern_das84}) and the equation simplifies. 
In \cite{ortalano97} this
equation was solved numerically for various geometries and the results
presented in the form of tables of pseudopotentials for quantum wells
and heterostructures for various values of the electron density and
device parameters. Here we show that the values of the pseudopotentials
characterizing the Coulomb interaction in the finite-width geometries
can be very accurately computed using a Gaussian and a new trial wf,
the `z\( \times  \)Gaussian' (zG), thereby allowing the effect on
the pseudopotentials of a finite-width in the direction perpendicular
to the 2D electron gas to be encoded in just one variational parameter,
\emph{ie} the width of the wf, parametrized by the standard deviation
\( w \) of the probability density. 

The self-consistent computation of \( \zeta (z) \) and \( V_{eff}(z) \)
is standard and follows the procedure given in \cite{ando_fs82,ortalano97}.
The potential \( V_{eff}(z) \) is written \begin{equation}
\label{V_(EFF) definition}
V_{eff}(z)=V_{W}(z)+V_{H}(z)+V_{XC}(z)
\end{equation}
 where \( V_{W}(z) \) is the confining potential of the quantum well
or heterostructure (including image charge effects and the effect
of the depletion layer) and \( V_{XC} \) is the exchange-correlation
potential \begin{equation}
\label{V_{XC}}
V_{XC}=\left[ 1+0.7734x\ln \left( 1+x^{-1}\right) \right] \left( \frac{2}{\pi \beta r_{s}}\right) R
\end{equation}
where \( \beta =(4/9\pi )^{1/3} \), \( x=r_{s}/21 \), \( r_{s}=(4\pi a^{*}n(z)/3)^{-1/3} \),
with \( a^{*} \)and \( R \) the effective Bohr radius and Rydberg
in GaAs. The Hartree potential is given by\begin{equation}
\label{Hartree potential}
V_{H}(z)=\frac{2\pi e^{2}}{\varepsilon }\int dz\int dz'|z-z'|\left( |\zeta (z)|^{2}-\rho (z)\right) \left( |\zeta (z')|^{2}-\rho (z')\right) 
\end{equation}
where \( \rho (z) \) is the (neutralizing) charge density of the
doping ions which are taken to be far way from the interface. In \cite{ando_fs82}
\( V_{H} \) is referred to as the potential due to the induced charges
or \( V_{S} \). In the presence of \( N_{A} \) acceptors per unit
volume in the semiconductor there will be \( n_{depl} \) (\( =N_{A}z_{D} \))
charges per unit area of the interface distributed evenly across the
depletion layer of width \( z_{D} \). 

We obtain \( \zeta (z) \) by solving (\ref{Schrodinger for zeta(z)})
using trial forms for \( \zeta (z) \) and compare the results with
those obtained by numerical solution in \cite{ortalano97}. The trial
waveforms we have studied are the Fang-Howard (FH), which is zero
for negative \( z \) and for positive \( z \) is given by \begin{equation}
\label{Fang-Howard}
\zeta (z)\propto z\exp (-bz/2),
\end{equation}
 the Gaussian \begin{equation}
\label{Gaussian wavefunction}
\zeta (z)\propto \exp (-(z-\alpha w)^{2}/4w^{2}),
\end{equation}
and the `z\( \times  \)Gaussian' (zG) wf, which is again zero for
negative \( z \) and for positive \( z \) is given by\begin{equation}
\label{z times Gaussian}
\zeta (z)\propto z\exp (-z^{2}/9c^{2}).
\end{equation}
 The width \( W \) of these wave functions can be characterized by
the standard deviation of the corresponding probability density. It
is given in terms of the parameters \( b,w,c \) as follows,

\begin{equation}
\label{eq:width_FH}
W_{FH}=\frac{\sqrt{3}}{b}
\end{equation}
for the Fang-Howard wf,

\begin{equation}
\label{eq:width_Gauss}
W_{G}=w
\end{equation}
for the Gaussian wf and

\begin{equation}
\label{eq:width_zGauss}
W_{zG}=\frac{3\sqrt{8+3\pi -16}}{2\sqrt{\pi }}c\approx 1.01016\times c
\end{equation}
for the zG wf. 

We determine the parameters \( b \), \( w \), \( \alpha  \) and
\( c \) variationally. We have found that, expanding the expression
for \( V_{XC} \) in (\ref{V_{XC}}) in \( x \) and keeping only
the constant and linear terms, reproduces the correct expectation
value for the exchange-correlation energy to within 0.1\% for all
three trial wf's, while including the quadratic term affects only
the fifth significant figure. The small \( x \) expansion works well
because the dominant contribution to the exchange-correlation energy
comes from the region in which the density is high (\( x \) small).
Using this expansion allows us, for the three trial forms for \( \zeta (z) \),
to compute analytically all the integrals involved in computing \( V_{eff}(z) \)
and, hence, also the expectation value for the sub-band energy \( E \).
For the case of the Gaussian trial wf, the effective interaction (\ref{2D from 3D interaction})
can be written in closed form in terms of the Bessel function \( K_{0}, \)\begin{equation}
\label{v_Gauss}
V_{G}(r)=\frac{1}{2\sqrt{\pi w^{2}}}e^{r^{2}/8w^{2}}K_{0}(r^{2}/8w^{2}).
\end{equation}

\begin{figure}
{\centering \resizebox*{7.8cm}{!}{\includegraphics{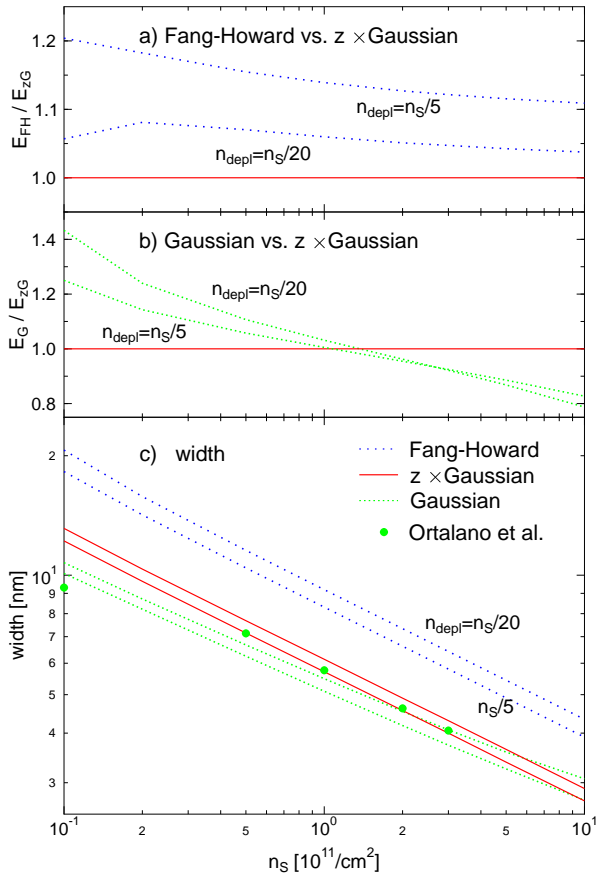}} \par}

\caption{\label{Trial wavefunctions compared}Comparison of the variational
estimates for the sub-band energies (see \ref{Schrodinger for zeta(z)})
and of the standard deviation of the charge distribution (width) for
the three variational wf's. The results for the width from direct
numerical solution of the equation by Ortalano et al \cite{ortalano97}
are also included. The top panel shows the ratio of the variational
estimates \protect\( E(\mbox {FH})/E(\mbox {zG})\protect \) and the
second panel shows \protect\( E(\mbox {Gauss})/E(\mbox {zG})\protect \).
The depletion layer density, \protect\( n_{depl}\protect \), is quoted
as a fraction of the electron density in the subband. }
\end{figure}

In Figure \ref{Trial wavefunctions compared} we compare the estimates
(see \ref{Schrodinger for zeta(z)}) of the sub-band energy for the
three variational wf's: \( E(\mbox {zG}) \), \( E(\mbox {Gauss}) \)
and \( E(\mbox {FH}) \) . For higher densities (\( n_{S}\gtrsim 10^{11} \)/cm\( ^{2} \)),
\( E(\mbox {Gauss}) \) gives the lowest variational estimate, while
for lower densities \( E(\mbox {zG}) \) gives the lowest estimate.
For all densities in the range we have studied, we find that the Fang-Howard
wf is worse as a variational wf than the zG and significantly worse
at higher densities than the Gaussian. This is because the FH wf has
too high a kinetic energy which it can only reduce by spreading the
density wider. Although the variational estimate of the energy for
the FH wf differs by a factor which only varies between 5\% and 20\%,
the width of its distribution, as measured by the standard deviation,
is significantly larger (\( \sim  \) 50\%) than for the other two
wf's. 

Given that the integrals involved in using the Gaussian or zG wf's
can be performed analytically and are more accurate as trial wf's,
it is perhaps surprising that these wf's have not been more widely
used in the study of heterostructures and quantum wells. Of the two,
the Gaussian is easier to use, although it will be less well-adapted
to MOS devices with large band gap discontinuities. For the heterostructures
considered below we use a (conduction) band gap discontinuity of 200
mV - the value appropriate for a GaAs/GaAs\( _{0.66} \)Al\( _{0.33} \)
heterostructure. On the other hand, the `zG', should become more favorable
as a variational wf, when the band gap discontinuity is large and
the effect of the boundary is well approximated by a hard wall.

\begin{figure}
{\centering \resizebox*{8.4cm}{!}{\includegraphics{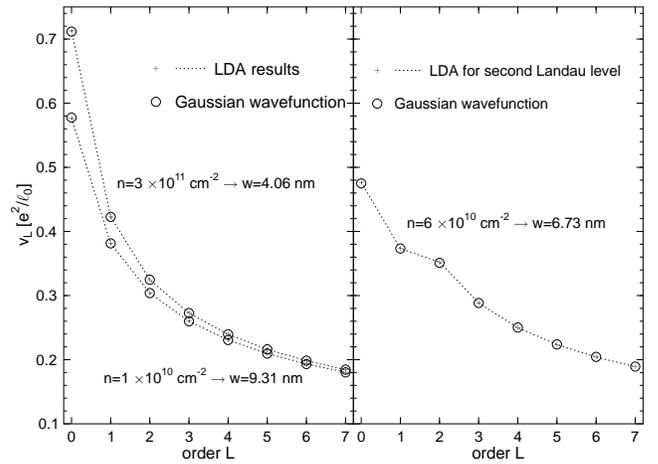}} \par}

\caption{\label{V(L) heterostructure/gaussian}The Haldane pseudopotentials
for the interaction \protect\( V_{G}(|\vec{r}_{1}-\vec{r}_{2}|)\protect \)
(\ref{v_Gauss}) projected onto the lowest and second Landau levels
as a function of angular momentum L for particles in heterostructures
with densities n. The results of direct numerical solution of equation
(\ref{Schrodinger for zeta(z)}) taken from \cite{ortalano97} are
shown as crosses and results based on the Gaussian trial wf for \protect\( \zeta (z)\protect \)
as circles. The width parameter w of (\ref{Gaussian wavefunction})
for the trial wavefunction is chosen so that the Haldane pseudopotential
\protect\( V_{1}\protect \) matches that obtained by numerical integration.
The differences between the results based on the Gaussian interface
wave function and the numerically computed local density approximation
is at the fraction of a percent level.}
\end{figure}

Figure \ref{V(L) heterostructure/gaussian} shows the Haldane pseudopotentials
for the interaction \( V(|\vec{r}_{1}-\vec{r}_{2}|) \) projected
onto the lowest and second Landau levels for heterostructures with
densities appropriate to samples studied experimentally. Here, we
determine the width of the trial interfacial wf's by requiring that
the Haldane pseudopotential \( V_{1} \) from \cite{ortalano97} is
exactly reproduced. We note here, that results for the pseudopotentials
in \cite{ortalano97} were for a value of the magnetic length \( \ell _{0} \)
which coincides with the Bohr radius \( a_{B}^{*}\approx 10nm \)
of electrons in GaAs. The results for the second Landau level were
for the density \( n_{s}=6\times 10^{10}cm^{-2} \) used in \cite{ortalano97}.
The study of the second Landau level in \cite{ortalano97} was motivated
by the results reported in \cite{willett_52} at filling fraction
\( \nu =5/2 \). However, the interfacial wave function is determined
by the \emph{total} number of electrons, which for the sample studied
in \cite{willett_52} was \( n_{s}\approx 3\times 10^{11}cm^{-2} \),
and not by the fraction occupying the second LL (\( n^{(1)}_{s}=6\times 10^{10}cm^{-2} \))
incorrectly used in \cite{ortalano97}. For this reason, the conclusions
regarding the \( \nu =5/2 \) state of Reference \cite{ortalano97}
are incorrect.

\begin{figure}
{\centering \resizebox*{8.2cm}{!}{\includegraphics{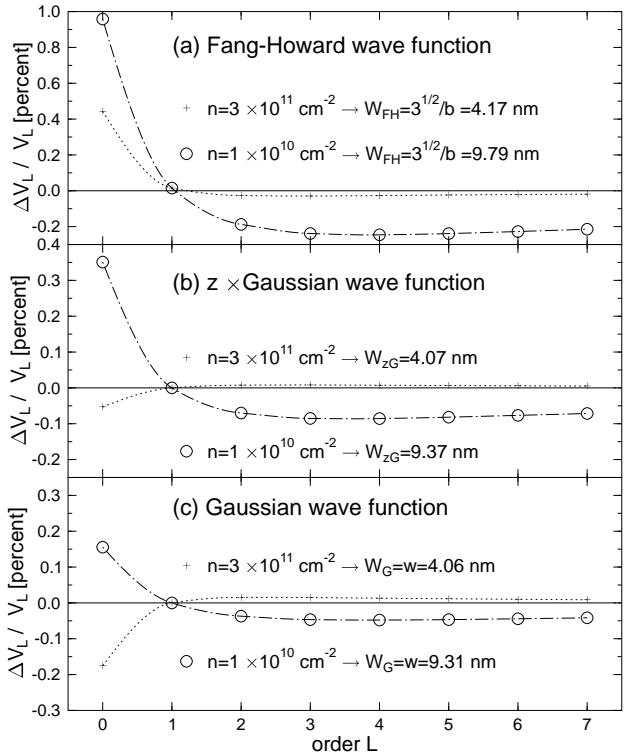}} \par}

\caption{\label{Delta V(L)}Errors in the Haldane pseudopotentials computed
using the Fang-Howard, z\protect\( \times \protect \)Gaussian (zG)
and Gaussian interface trial wf's for two different densities. The
comparison is with the values reported in \cite{ortalano97}. The
variational parameters are determined such that \protect\( V_{1}\protect \)
is correctly reproduced. The zG leads to errors for \protect\( V_{m},\, m\neq 1\protect \)
which are roughly half as big as those with the Fang-Howard wf. The
Gaussian wf is even better in reproducing the LDA results with a maximum
error of less then 0.2 percent (for \protect\( V_{0}\protect \)).
The width parameters for the three variational wavefunctions (given
above each curve) are approximately equal for both densities for all
three wavefunctions, implying that fixing the pseudopotential \protect\( V_{1}\protect \)
is effectively equivalent to fixing the standard deviation of the
charge distribution.}
\end{figure}

It is clear that the use of the Gaussian trial wf yields results which
are essentially indistinguishable from the results of the exact numerical
solution for \( \zeta (z) \). We find very similar results for the
zG. In Figure \ref{Delta V(L)} we show the difference between the
pseudopotentials computed exactly by solving numerically for the interface
wf \( \zeta (z) \) (taken from \cite{ortalano97}) and those obtained
using the Gaussian and FH wf's. The errors obtained using the FH wf
are at the 1\% level while those obtained for the Gaussian are at
the 0.1\% level. Those obtained using the Gaussian trial wf are smaller
than other uncertainties in the model such as those related to the
value chosen for the depletion density, \( n_{depl} \). Finite-width
effects on the Haldane pseudopotentials are clearly accurately captured
by the Gaussian (and zG) trial wf's. Given the fact that the pseudopotentials
\( V_{m} \) only depend on the width parameter (\( w \) for the
Gaussian wf, \( b \) for the Fang-Howard wf and \( c \) for the
zG), it is clear that the use of these trial wf's massively simplifies
the study of finite-width effects when compared to the numerical integration
of (\ref{Schrodinger for zeta(z)}) and tabulation of pseudopotentials
used in \cite{ortalano97}. For the case of the Gaussian, we also
have an analytic expression for the effective interaction \( V(|\vec{r}_{1}-\vec{r}_{2}|) \),
cf equation (\ref{v_Gauss}).

The tables I-IV of reference \cite{ortalano97} can be summarized
by listing the effective width of the Gaussian interface wf for which
the Haldane pseudopotential of order \( m=1 \) is exactly reproduced.
In Table \ref{ortalano_table} we list the width parameters for all
tabulated cases.

\begin{table}
\begin{tabular}{|c|c|c|c|}
\hline 
I&
II&
III&
IV\\
parabolic QW&
Heterointerface&
square QW&
Heterointerface\\
\hline
\( n_{s}\, \, \, \, \, \, \, \, \, \, \, \, \, w[nm] \)&
\( n_{s}\, \, \, \, \, \, \, \, \, \, \, \, \, w[nm] \)&
\( n_{s}\, \, \, \, \, \, \, \, \, \, \, \, \, w[nm] \)&
\( n_{s}\, \, \, \, \, \, \, \, \, \, \, \, \, w[nm] \)\\
0.49~~~~~19.3813&
0.10~~~~~9.30690&
0.10~~~~~3.21628&
0.60~~~~~6.72784\\
0.60~~~~~24.1478&
0.50~~~~~7.13556&
0.50~~~~~3.21854&
\\
0.73~~~~~28.9638&
1.00~~~~~5.76001&
1.00~~~~~3.22610&
\\
0.85~~~~~33.1890&
2.00~~~~~4.61801&
5.00~~~~~3.31037&
\\
&
3.00~~~~~4.06015&
&
\\
\hline
\end{tabular}

\caption{\label{ortalano_table}Width parameters for all tabulated results
of Reference \cite{ortalano97}. The electron density \protect\( n_{s}\protect \)
is measured in units of \protect\( 10^{11}/cm^{2}\protect \) while
the width \protect\( w\protect \) of the interfacial wf is given
in \protect\( nm\protect \). }
\end{table}

\begin{figure}
{\centering \resizebox*{7.8cm}{!}{\includegraphics{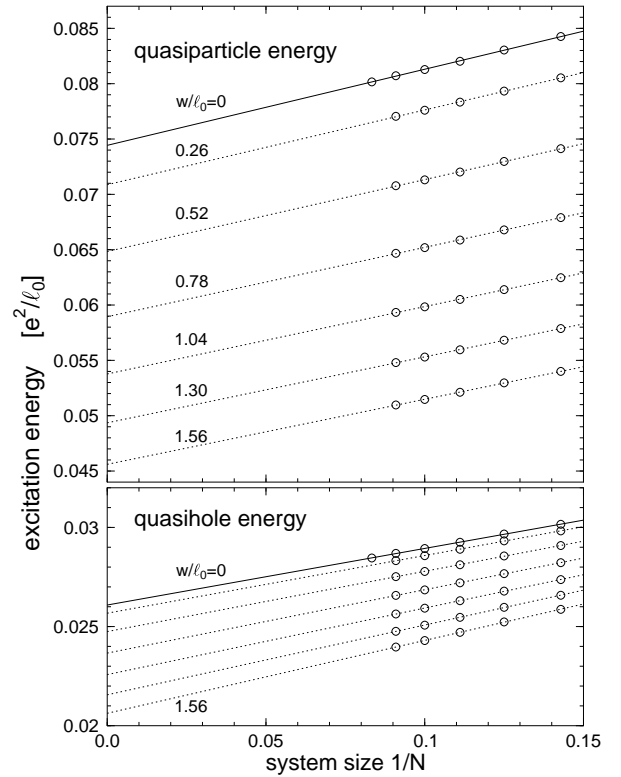}} \par}

\caption{\label{qpqh1/3(width)}The quasihole energy \protect\( \widetilde{e}^{+}_{1/3}(N)\protect \)
and quasiparticle energy \protect\( \widetilde{e}^{-}_{1/3}(N)\protect \)
and the best linear fits to these points computed as a function of
the width \protect\( w\protect \) of the density distribution computed
using Gaussian trial wf's. The sum of the two linear functions can
be taken as a measure of the gap energies \protect\( \widetilde{e}^{g}_{1/3}(N)\protect \). }
\end{figure}

\section{Finite Width Effects on Energy Gaps}

\subsection{Filling fractions \protect\( \nu =\frac{p}{2p+1}\protect \)}

\begin{figure}
{\centering \resizebox*{8cm}{!}{\includegraphics{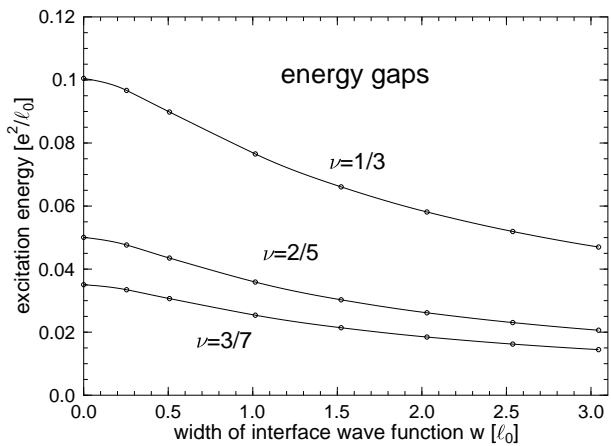}} \par}

\caption{\label{gaps(width)}Estimates of the energy gaps in the thermodynamic
limit as a function of the width of the subband wf (taken as the standard
deviation of the charge distribution). The solid lines show the fits
to the interpolation formula (cf. Equation (\ref{eq:interpol}) and
Table \ref{table:interpol}).}
\end{figure}

With these modified potentials we have repeated the calculations described
in section 2. Using the Gaussian wf parametrized by its width \( w \)
we compute the Haldane pseudopotentials as a function of \( w \).
By exact diagonalization just as in the pure Coulomb case of Section
3, we compute width dependent excitation energies for all possible
system sizes and perform for each value of the width parameter \( w \)
an extrapolation to the thermodynamic limit \( N\rightarrow \infty  \).
As an example we show in Figure \ref{qpqh1/3(width)} the size and
width dependent quasiparticle \( \widetilde{e}^{-}_{1/3}(N) \) and
quasihole energies \( \widetilde{e}^{+}_{1/3}(N) \) at \( \nu =1/3 \).
For each width \( w \) we use linear extrapolation in \( 1/N \)
to estimate the gap energy in the thermodynamic limit as a function
of width. The size dependence at finite width is qualitatively the
same as at \( w=0 \). This same procedure was also employed for the
calculation of width dependent quasiparticle and quasihole energies
at \( \nu =1/3 \) and \( \nu =2/5 \), and the corresponding energy
gaps in the thermodynamic limit. The result of these calculation are
shown in Figure \ref{gaps(width)}. The full lines in Figure \ref{gaps(width)}
correspond to interpolation functions of the form

\begin{equation}
\label{eq:interpol}
E_{G}(x)=E^{(0)}_{G}\times (\frac{\cos ^{2}\phi }{\sqrt{1+ax^{2}}}+\frac{\sin ^{2}\phi }{1+bx^{2}}),
\end{equation}
where \( x=w/\ell _{0} \). The functional form (\ref{eq:interpol})
is suggested by the following observations: The Haldane pseudopotentials
\( V_{m} \) for \( m>0 \) behave for \( w\rightarrow 0 \) as \( V_{m}\approx V^{(0)}_{m}+O(w^{2}) \)
while for very large \( w \) they behave as \( V_{m}\approx (\log (w)/\sqrt{\pi }+\alpha _{m})/w \).
Indeed, as will be seen below, the energy gaps decrease as \( 1/w \)
for very large values of the width \( w \), implying that the logarithmic
term cancels out in this limit. The values of the fitting parameters
\( E^{(0)}_{G},\, \phi ,\, a \) and \( b \) are listed in Table
\ref{table:interpol} for the filling fractions \( \nu =\frac{1}{3},\, \frac{2}{5},\, \frac{3}{7} \). 

\begin{table}
\begin{tabular}{|c|c|c|c|c|}
\hline 
\( \, \, \nu \, \,  \)&
\( E^{(0)}_{G}\, \,  \){[}\( \frac{e^{2}}{\epsilon \ell _{0}} \){]}&
\( \phi  \)~{[}degrees{]}&
\( a \)&
\( b \)\\
\hline
1/3&
0.1012&
34.18&
0.1468&
1.542\\
2/5&
0.0500&
36.07&
0.1935&
1.866\\
3/7&
0.0350&
34.98&
0.2078&
1.851\\
\hline
\end{tabular}

\caption{\label{table:interpol}Interpolation function for the gap energy
as a function of width (\ref{eq:interpol}): parameters for filling
fractions \protect\( \nu =1/3,\, \, 2/5,\, \, 3/7\protect \). }
\end{table}

The results presented in Figure \ref{gaps(width)} are similar to
those reported in \cite{park_mj99}. The results of \cite{park_mj99}
were based on Monte Carlo simulations (MC) of CF trial wavefunctions,
which as mentioned in section IV, give larger gaps than our results
for the bare Coulomb interaction even at \( \nu =1/3. \) This discrepancy
exists throughout the range of \( w/\ell _{0} \) in the figure with
our estimates being between \( \sim  \)5\% smaller (for \( \nu =1/3 \))
and \( \sim 25 \)\% smaller (for \( \nu =3/7 \) and \( 4/9 \)).
(It is not surprising that the difference does not depend strongly
on \( w/\ell _{0} \): While the energies are affected by the width
\( w \) through the variation of the effective interaction, the wavefunctions
are not expected to change significantly \cite{haldane_reza85}.)
The gaps as a function of width have also been estimated \cite{murthy99}
using a field theoretic approach \cite{murthy99}, which constructs
explicit CF wavefunctions out of Chern-Simons gauge-transformed fermions.
Energies of ground and excited states can be computed within this
theory at the Hartree-Fock level. The theory needs to cut off the
interaction at large wavevectors and should therefore be reliable
for large widths where the inverse width provides a natural large
wavevector cut-off. For widths \( w/\ell _{0}\gtrsim 2 \), the results
are consistent with those of the MC simulations using composite fermion
trial wf's \cite{park_mj99}. For \( 0.5<w/\ell _{0}<2 \), the results
are still comparable to those of the MC simulations, although they
imply gaps which are rapidly increasing as \( w/\ell _{0}\rightarrow 0 \),
in contrast to the results in Figure \ref{gaps(width)}.

\begin{figure}
{\centering \resizebox*{7.8cm}{!}{\includegraphics{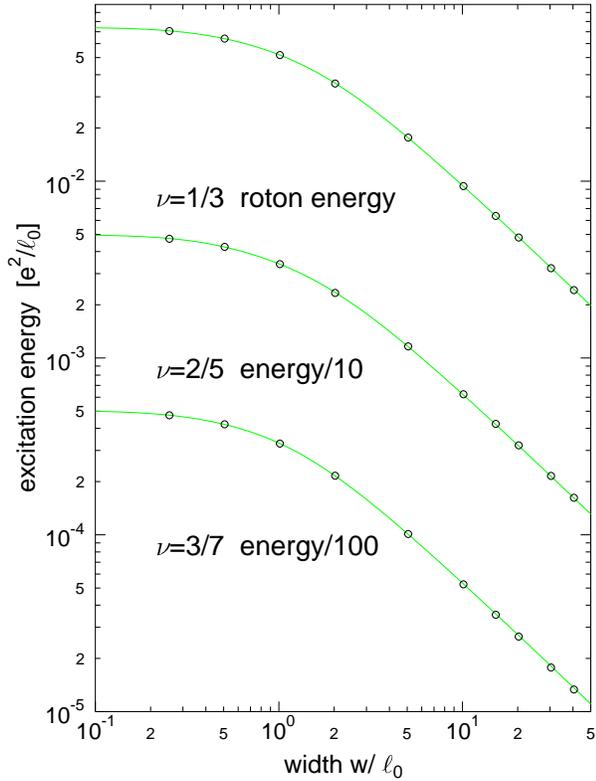}} \par}

\caption{\label{rotons_at_large_widths}The energy of the roton minimum as
a function of width for systems with 11 particles at \protect\( \nu =1/3\protect \),
14 particles at \protect\( 2/5\protect \) and 15 particles at \protect\( 3/7\protect \).
For clarity, the energies of the roton are scaled by a factor 1/10
at \protect\( \nu =2/5\protect \) and 1/100 at \protect\( \nu =3/7.\protect \)
If there were to be phase transition to a compressible state the gap
would have to vanish at some finite width. Instead we find clear evidence
that, for large widths, the energy of the roton minimum (the lowest-lying
excitation at fixed total flux) tends to zero as \protect\( 1/w\protect \)
with no suggestion of a phase transition.}
\end{figure}
Evidence, also based on CF trial wavefunctions, was presented in \cite{park_j99}
which suggested a phase transition as a function of increasing width
from incompressible states to compressible states at filling fractions
\( \nu =p/(2p+1) \). We have tested this theoretical prediction by
computing the width dependent energy of the lowest energy excitation,
which corresponds to the roton minimum. We have not analyzed the extrapolation
to the infinite-system size limit for the roton minimum and present,
instead, the variation with width of the roton minimum energy for
a system with fixed particle number. We show the results for \( \nu =1/3 \),
\( 2/5 \) and \( 3/7 \) in Figure \ref{rotons_at_large_widths}.
We find that even for very large and unphysical widths up to hundreds
of nanometers (corresponding to up to \( 50\, \ell _{0} \)), there
is no evidence of the gap vanishing at any of these filling fractions.
Instead we find that for such large width parameters the roton minimum
scales simply as \( 1/w \).

\subsection{\protect\( \nu =\frac{5}{2}\protect \) state}

Here, we present the results of calculations of the finite width effect
on the energy gap of the mysterious \( \nu =5/2 \) state. If the
effects of Landau level mixing are neglected, it is sufficient to
solve for the ground state of the electrons occupying the second -
half-filled - Landau level and take the filled lowest Landau levels
of spin-up and spin-down electrons as inert, i.e. unpolarizable. This
problem is characterized by a filling factor \( \nu ^{(1)} \) of
the first excited Landau level of \( \nu ^{(1)}=1/2. \) It is customary
to represent the system of electrons filling half the second Landau
level by lowest Landau level wave functions but to take into account
the interaction of electrons in the second Landau level by using the
appropriate Haldane pseudopotentials of the second Landau level. Again,
as for the computation of energy gaps at \( \nu =p/(2p+1) \), there
are essentially two ways to compute the energy. Either one may calculate
neutral excitation (exciton) energies corresponding to a widely separated
quasiparticle and quasihole pair, or one may calculate the energy
of ground states containing a (fractionally) charged excitation. In
the case of the \( \nu =5/2 \) state, or equivalently at \( \nu ^{(1)}=1/2 \),
there is the problem that elementary charged excitations are predicted
to occur only in pairs.

\begin{figure}
{\centering \resizebox*{8cm}{!}{\includegraphics{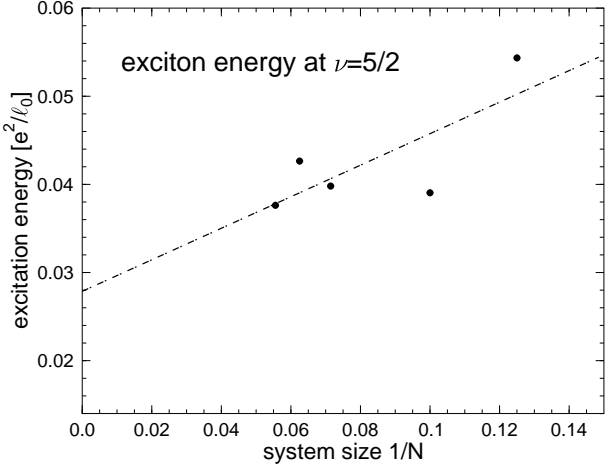}} \par}

\caption{\label{fig:exciton52}The exciton energy at \protect\( \nu =5/2\protect \)
for zero width, corrected for the Coulomb attraction between quasiparticle
and quasihole (\protect\( A_{1/4}(1/2)\protect \), equation (\ref{eq:aq}),
is plotted as a function of system size \protect\( 1/N\protect \).}
\end{figure}

The polarized ground state at \( \nu ^{(1)}=1/2 \) occurs on the
sphere when the number of flux units is 

\begin{equation}
\label{eq:2S_52}
2S_{0}=2N-3,
\end{equation}
and is thought to be described by a paired state, which may be of
the Moore-Read pfaffian type \cite{Moore-Read,morf98,Rez-Hald00}.
However, great care is needed when analyzing excitation energies in
these states on the sphere to avoid mistaking systems at conventional
filling fractions \( \nu _{p}=p/(2p+1) \) or \( 1-\nu _{p} \) for
systems at filling \( \nu ^{(1)}=1/2. \) As we have discussed previously
\cite{nda_morf89}, systems on the surface of a sphere exhibit degeneracies
where, for a certain size \( N \), states with different filling
factor coincide. This turns out to be a particularly severe problem
in the sequence (\ref{eq:2S_52}). Indeed, of the possible systems
with up to 18 electrons, only five are not aliased with conventional
fractional states, namely those with \( N=8,10,14,16 \) and 18 particles.
Of these, the ones at \( N=8 \) and 16 have the problem that charged
excitations of these states are aliased with ground states of conventional
FQH states. Using these aliased states for a calculation of the energy
gap at \( \nu ^{(1)}=1/2 \) would be misleading and would give rise
to systematic errors. In Table \ref{tab:aliases_52}, we list the
relevant states and their aliases.
\begin{table}
\begin{tabular}{|c|c|c|c|c|c|c|}
\hline 
\( N \)&
\( 2S_{0} \) (GS)&
\( \nu _{a} \)&
\( 2S_{0}+1 \)&
\( \nu _{a} \)&
\( 2S_{0}-1 \)&
\( \nu _{a} \)\\
\hline
6&
9&
2/3&
10&
&
8&
\\
8&
13&
&
14&
&
12&
2/3\\
10&
17&
&
18&
&
16&
\\
12&
21&
3/5&
22&
&
20&
\\
14&
25&
&
26&
&
24&
\\
16&
29&
&
30&
4/5&
28&
\\
18&
33&
&
34&
&
32&
\\
\hline
\end{tabular}

\caption{\label{tab:aliases_52}Total flux in the ground (\protect\( 2S_{0}\protect \))
and excited (\protect\( 2S_{0}\pm 1\protect \)) states for systems
at \protect\( \nu ^{(1)}=1/2\protect \) as a function of number of
particles, \protect\( N\protect \). Where these states are aliased
to conventional quantum Hall state ground states, we also show the
corresponding filling fractions, \protect\( \nu _{a}\protect \).
We note that the only sizes for which no aliases occur, are \protect\( N=10,14\protect \)
and 18. Unaliased ground states occur in addition at \protect\( N=8\protect \)
and 16.}
\end{table}
 We first show the energy of neutral excitations (exciton) with maximal
angular momentum \( L_{max} \), corresponding to the largest possible
separation of the quasiparticle and quasihole on the sphere. The angular
momentum of this exciton is given by \( L_{max}=N/2 \) if \( N/2 \)
is even, otherwise \( L_{max}=N/2-1 \). In Figure \ref{fig:exciton52}
the exciton energy for zero width, corrected for the Coulomb attraction
between quasiparticle and quasihole (\( A_{1/4}(1/2) \), equation
(\ref{eq:aq}), is plotted as a function of system size \( 1/N \)
together with a linear fit in \( 1/N \) to the data at \( N=8,10,14,16 \)
and 18, cf. Table \ref{tab:aliases_52}. Like at \( \nu =2/5, \)
the exciton energy shows very large, and fluctuating finite size effects.
Extrapolation to the bulk limit using a linear fit in \( 1/N \) yields
the result

\begin{equation}
\label{eq:exciton52}
\Delta ^{exc}_{5/2}\approx 0.028\frac{e^{2}}{\epsilon \ell _{0}}.
\end{equation}

Alternatively, the energy gap can also be computed by calculating
individually the energy of quasiparticle and quasihole excitations.
The two quasiparticle state occurs at \( 2S_{0}-1 \) while the two
quasihole state occurs for \( 2S_{0}+1 \). Since in both cases the
two excitations have the same charge (\( q=e/4 \) for the quasiparticle
and \( q=-e/4 \) for the quasihole), one expects that the lowest
energy state occurs when the two charges are maximally far apart,
which demands maximum relative angular momentum, and consequently
minimum total angular momentum on the sphere. Although one might have
expected that this would imply \( L=0 \) for the ground state, as
a result of symmetry, the angular momentum of the lowest energy states
is \( L=N/2\, mod\, 2, \) i.e. \( L=1 \) for \( N=10,14,18. \)
The energy of these two-quasiparticle or two-quasihole states contains,
in addition to the term \( A_{2q}(\nu ^{(1)}) \) (equation \ref{eq:aq}),
the Coulomb interaction, \( \Delta A_{q}, \) of two quasiparticles
separated by twice the radius \( R \) (the maximal separation on
the sphere):

\begin{equation}
\label{eq:delta_aq}
\Delta A_{q}=q^{2}\sqrt{\frac{\nu ^{(1)}}{2N}}.
\end{equation}
Combining the two contributions \( A_{2q}(\nu ^{(1)}) \) and \( \Delta A_{q} \)
gives for the finite size correction term \( C_{q}(N) \)

\begin{equation}
\label{eq:Cq(N)}
C_{q}(N)=-3q^{2}\sqrt{\frac{\nu ^{(1)}}{2N}}=-\frac{3}{32}\sqrt{\frac{1}{N}}.
\end{equation}
The gap calculation then proceeds by taking account explicitly of
the finite size correction \( C_{q}(N) \) (equation \ref{eq:Cq(N)}),
as described for the cases at \( \nu =p/(2p+1) \) in the previous
section.

\begin{figure}
{\centering \resizebox*{8cm}{!}{\includegraphics{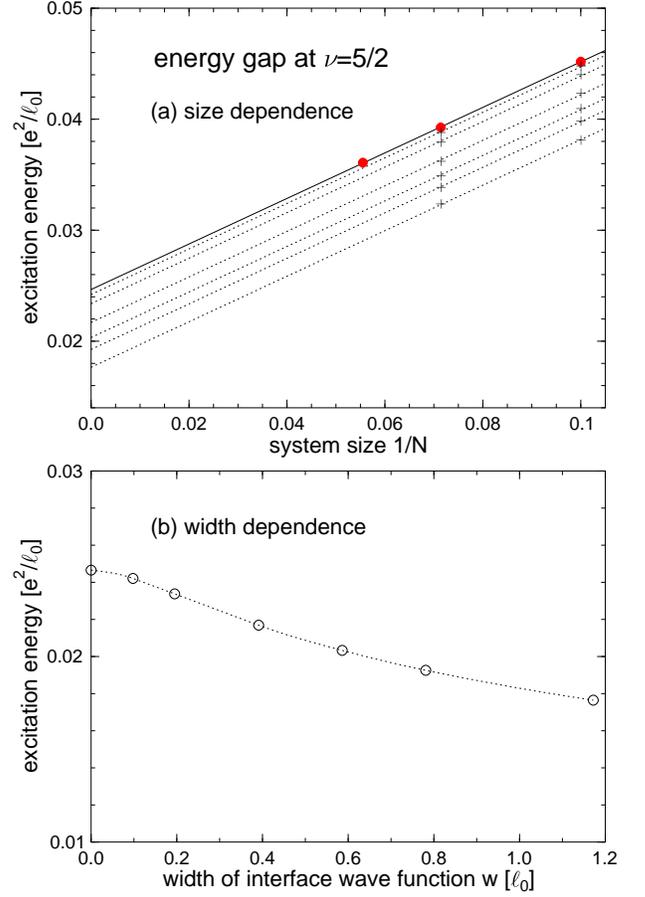}} \par}

\caption{\label{fig:nu52}Energy gap at \protect\( \nu =5/2.\protect \) The
upper panel illustrates the size dependence of the energy gap for
different values of the width parameter \protect\( 0\leq w/\ell _{0}\leq 1.17.\protect \)
The width parameters are \protect\( w/\ell _{0}=0,\, 0.098,\, 0.195,\, 0.391,\, 0.586,\, 0.781,\, 1.17,\protect \)with
the topmost line referring to the case \protect\( w=0\protect \),
and the rest in sequence down to the lowermost line with \protect\( w/\ell _{0}=1.17\protect \).
The extrapolations to the \protect\( N\rightarrow \infty \protect \)
limit assume that the slopes for the cases with \protect\( w\neq 0\protect \)
are the same as for the \protect\( w=0\protect \) case. In the lower
panel the gap values, extrapolated to the \protect\( N\rightarrow \infty \protect \),
are plotted as a function of \protect\( w/\ell _{0}\protect \).}
\end{figure}

In Figure \ref{fig:nu52}(a), we show our results for the gap at \( \nu =5/2 \).
In the top figure, half the sum of quasiparticle and quasihole excitation
energies are plotted as a function of system size \( 1/N \) for different
values of the width \( w \). For zero width, results for \( N=10,14 \)
and 18 are plotted, the sizes at which no aliasing effects occur.
They lie almost exactly on a straight line in \( 1/N. \) Extrapolation
to the bulk limit yields

\begin{equation}
\label{eq:gap52}
\Delta _{5/2}\approx 0.025\frac{e^{2}}{\epsilon \ell _{0}},
\end{equation}
slightly lower, but consistent with the result (\ref{eq:exciton52})
based on the exciton energies. Based on our previous experience with
gap calculations at \( \nu =1/3 \) and 2/5, we believe that also
at \( \nu =5/2 \) the extrapolation based on individual quasiparticle
and quasihole energies is more reliable. However, the exciton energy
calculation certainly supports our conclusion that the quasiparticle
and quasihole states at \( \nu =5/2 \) contain two charged defects.
Otherwise, there would be a factor of two difference between our extrapolated
values \( \Delta ^{exc}_{5/2} \) (equation \ref{eq:exciton52}) and
\( \Delta _{5/2} \) (equation \ref{eq:gap52}). Finally, in Figure
\ref{fig:nu52}(b), the gap in the thermodynamic limit is plotted
as a function of width \( w/\ell _{0} \). These results indicate
that the width effects reduce the gap at \( \nu =5/2 \) slightly.

Very recently, Eisenstein et al \cite{eisenstein_2002}  have investigated the 
$\nu =5/2$ and 7/2 states in a sample of ultra-high mobility 
($\mu \approx 3.1\times 10^{7}$cm$^{2}$/Vs). They determined an 
activation gap $\Delta ^{m}_{5/2}\approx 0.31$K at $\nu =5/2$ and 
$\Delta ^{m}_{7/2}\approx 0.07$K
at $\nu =7/2$. Their sample had an electron density $n_{s}=3\times 10^{11}$/cm$^{2}$, 
which leads to a width $w\approx 4$nm. At $\nu =5/2$, the field $B=4.96$T 
corresponds to a value  $w/\ell _{0}\approx 0.35$, while at $\nu =7/2$ 
we get $w/\ell _{0}\approx 0.30$. 
According to the results shown in Figure \ref{fig:nu52}, 
The calculated gap values (see Figure \ref{fig:nu52}) for $w/\ell _{0}\approx 0.35$ 
and 0.30 are  around 0.0220 and  0.0225$e^{2}/\epsilon \ell _{0}$ respectively. 
These lead to theoretical values for the gap of $\Delta ^{m}_{5/2}\approx 2.5$K and 
$\Delta ^{m}_{7/2}\approx 2.1$K. A disorder broadening of the order of 2K would 
explain the measured gaps of 0.31K and 0.07K. It is important 
to note that previous experimental values of the excitation gap at $\nu =5/2$
have been much smaller \cite{eisenstein_52_1,pan_exact_quantization}. 
For samples with density  $n_{s}=2.3\times 10^{11}$/cm$^{2}$ the gap 
at $\nu=5/2$ was 0.11K \cite{pan_exact_quantization}. 
In this case, the width is 
$w\approx 4.5$nm, and at the field $B=3.65$T, we obtain 
$w/\ell _{0}\approx 0.34$, and a theoretical gap value of 
2.1K. The factor of $\sim 3$ difference between the recently
reported activation gap \cite{eisenstein_2002} and the earlier estimate
\cite{pan_exact_quantization} in samples with
similar densities suggests that  
the activation gap  is affected strongly by sample quality, 
and is likely to be dominated by the effects of disorder. 
By comparing the 
gap at 5/2 to those at 7/3 and 8/3, and also at $\nu =p/(2p+1)$, Pan et al.  
also concluded that a disorder broadening of the order 2K was to be expected.


\section{Experimental Gaps}

Estimates of the gaps for fractional quantum Hall systems have been
reported for GaAs heterojunctions \cite{willett88,du93} and more
recently for metal-insulator-structures (MIS) using organic (pentacene
and tetracene) semiconductor layers \cite{tetracene_gaps}. The recent
measurements on organic MIS structures are particularly interesting
given the different separation of energy scales to that found in GaAs.
The dielectric constant in tetracene is in the range \( \epsilon \approx  \)3
to 4 (compared to \( \epsilon \approx  \)12.7 in GaAs), the band
mass is \( \sim 1.3 \)\( m_{e} \) (0.07 in GaAs) and the \( g \)-factor
is close to 2 (0.44 in GaAs). The larger band mass and the smaller
dielectric constant mean that, for samples with the same density,
the ratio of interaction energies to the Landau-level splitting is
much larger in the tetracene structures than in GaAs and hence that
Landau level mixing effects are expected to be larger. The larger
\( g \)-factor gives a larger Zeeman energy, and hence makes spin-reversed
excitations less likely than in GaAs heterostructures.

\begin{table}
\begin{tabular}{|c|c|c|c|c|c|c|c|c|}
\hline 
&
\multicolumn{4}{c|}{Sample A (\( \Gamma =1.28\pm .13 \)K)}&
\multicolumn{4}{c|}{Sample B (\( \Gamma =2.1\pm .17 \) K)}\\
\hline
\( \, \, \, \nu \, \, \,  \)&
B{[}T{]}&
\( \Delta ^{m}_{\nu } \) \( +\Gamma  \) {[}K{]}&
\( \Delta ^{c}_{\nu } \) {[}K{]}&
\( \frac{\Delta ^{m}_{\nu }+\Gamma }{\Delta ^{c}_{\nu }} \)&
B{[}T{]}&
\( \Delta ^{m}_{\nu } \) \( +\Gamma  \){[}K{]}&
\( \Delta ^{c}_{\nu } \){[}K{]}&
\( \frac{\Delta ^{m}_{\nu }+\Gamma }{\Delta ^{c}_{\nu }} \)\\
\hline 
1/3&
13.9&
9.03&
15.2&
0.59&
28.5&
13.2&
18.9&
0.70\\
2/5&
11.6&
4.48&
6.9&
0.65&
23.8&
6.5&
8.6&
0.75\\
3/7&
10.8&
3.23&
4.8&
0.67&
23.2&
4.5&
5.7&
0.79\\
4/9&
10.4&
2.23&
3.6&
0.61&
21.4&
3.3&
4.9&
0.68\\
\hline
\end{tabular}

\caption{\label{Du_A&B}Comparison of the measured gaps, \protect\( \Delta ^{m}_{\nu }\protect \),
in samples A (nominal density \protect\( 1.12\times 10^{11}\protect \)
cm\protect\( ^{-2}\protect \)) and B (nominal density \protect\( 2.3\times 10^{11}\protect \)cm\protect\( ^{-2}\protect \))
reported in \cite{du93} with the gaps computed for a Coulomb interaction
but taking account of the finite-width effects, \protect\( \Delta ^{C}_{\nu }\protect \).
We have added a constant field-independent shift, \protect\( \Gamma ,\protect \)
for each sample which we estimate by comparing the functional dependence
of the gap energies as a function of filling fraction, \protect\( \nu ,\protect \)
predicted by CF theory with that found in experiment. The range quoted
for \protect\( \Gamma \protect \) gives the maximum and minimum found
when the constant \protect\( C'\protect \) in (\ref{Gaps_in_CStheory})
varies between \protect\( 4.11\protect \) (our estimate of \protect\( C'\protect \)
for the pure Coulomb interaction) and \protect\( 9\protect \). }
\end{table}

We have estimated the gaps at \( \nu =1/3 \), \( 2/5 \) and \( 3/7 \)
for the two samples A and B of \cite{du93}. We take the quoted density
of the samples and assume a depletion density \( n_{depl}=n_{S}/5 \)
(this is typical of these samples \cite{willett88}, although the
results are not sensitive to the exact value of \( n_{depl} \)).
>From the results in Fig \ref{Trial wavefunctions compared} we estimate
the standard deviation of the density distribution and this leads
directly to an estimate of the gaps (see Fig \ref{gaps(width)}).
We compare our results with those of the two samples A and B of \cite{du93}
in Table \ref{Du_A&B}. The effects of impurity scattering have been
taken into account empirically by assuming that the levels are broadened
by a field-independent broadening \( \Gamma  \). This assumption
has not been theoretically justified, However, for the purpose of
comparison we have reanalyzed the results of \cite{du93} under this
assumption by fitting them to the functional form predicted by CF
theory, i.e. including the logarithmic corrections (see \ref{Gaps_in_CStheory})
to extract the broadening \( \Gamma  \). We find that the gaps measured
are consistently between 60 and 70\% of what we predict after taking
account of finite thickness effects. This is consistent with the results
of \cite{morf_park_j99,park_mj99}, correcting the error of Reference
\cite{park_j99}.

\begin{table}
\begin{tabular}{|c|c|c|c|c|}
\hline 
\( \, \, \, \nu \, \, \,  \)&
B{[}T{]}&
\( \Delta ^{m}_{\nu } \) {[}K{]}&
\( \Delta ^{c}_{\nu } \) {[}K{]}&
\( \Delta ^{w}_{\nu } \) {[}K{]}\\
\hline
1/3&
21.0&
10.5&
16.6&
13.5\( \pm .5 \)\\
2/3&
10.8&
6.5&
13.1&
10.7\( \pm .5 \)\\
5/3&
4.5&
1.0&
9.3&
7.9\( \pm .3 \)\\
\hline
\end{tabular}

\caption{\label{Willett_data}The activation energies as deduced from the
temperature dependence of the longitudinal resistivity at filling
fractions \protect\( \nu =1/3,\, 2/3,\, 5/3\protect \), \protect\( \Delta ^{m}_{\nu }\protect \),
reported in \cite{willett88} are compared to our values for the gaps,
\protect\( \Delta ^{c}_{\nu }\protect \). For reference, we also
show the calculated values of Willett et al. \cite{willett88} in
the last column. These authors fixed the depletion density \protect\( n_{depl}\protect \)
and hence the width parameter \protect\( w\protect \) by requiring
that the solution of (\ref{Schrodinger for zeta(z)}) correctly reproduced
the experimentally measured sub-band splitting. They estimated the
finite width corrections on the basis of the model interaction (\ref{Zhang_das_Sarma_Interaction})
to give \protect\( \Delta ^{w}_{\nu }\protect \) .}
\end{table}

The results reported in \cite{willett88} relate to filling fractions
\( p/3 \), where \( p=1 \), 2, 4 and 5 and were interpreted on the
assumption that the ground states and gaps were all maximally spin-polarized
states within the lowest Landau level for a sample with density \( n_{s}=1.65\times 10^{11} \)
cm\( ^{-2} \) and mobility \( 5\times 10^{6} \) cm\( ^{2} \)/V
sec (to be compared with \( 6.8 \) and \( 12\times 10^{6} \) cm\( ^{2} \)/V
sec in samples A and B in \cite{du93}). The authors of \cite{willett88}
solved (\ref{Schrodinger for zeta(z)}) numerically for the sub-band
wf, \( \zeta (z) \), choosing the depletion density \( n_{depl} \)
to reproduce the experimentally observed sub-band splitting. As as
a result we have a more precise estimate of width of the wf in the
lowest sub-band than we have been able to make for the samples of
\cite{du93}. We have converted their estimate of the width to a standard
deviation \( w \) and estimated the gaps at the relevant filling
fractions. The results are presented in Table \ref{Willett_data}.
We note that the measured values \( \Delta ^{m} \) of the gap at
\( \nu =1/3 \) and 2/3 are both larger than our theoretical values
by about the same amount \( \Gamma \approx 6 \)K. This might serve
as an estimate of the broadening. The authors of \cite{willett88}
also estimated the gap reduction on account of finite thickness effects
based on the exact diagonalizations of six particle systems reported
in (\ref{Zhang_das_Sarma_Interaction}) and we include these estimates
\( \Delta ^{w}_{\nu } \) in the Table. The reduction of the gaps
found in the earlier finite-size studies was significantly larger
than what we obtain (Section 3). It may have resulted from estimating
the gap reduction using systems which were too small, or inaccurate
extrapolation to the thermodynamic limit.

It is clear from both Tables \ref{Du_A&B} and \ref{Willett_data}
that the discrepancy between measured gaps and computed gaps is significant.
This discrepancy may be due to Landau level mixing, spin-reversed
excitations and to impurity effects not accounted for by the use of
the field-independent broadening \( \Gamma  \) used in Table \ref{Du_A&B}.
Estimates in \cite{yoshioka86} based on diagonalizations of up to
only 5 particles in a torus geometry suggested reductions of the gap
(identified with the zone boundary exciton) as a result of Landau
level mixing of between 12\% and 17\% were possible at \( \nu =1/3 \)
in a magnetic field at 10T for a pure Coulomb interaction. These should
scale as \( (e^{2}/\varepsilon \ell _{0})/\hbar \omega _{c}\sim 1/\sqrt{B} \).
On this basis the reduction at a field of 28.5T for sample B at \( \nu =1/3 \)
would be at most 10\%. However, as the matrix elements between Landau
levels of the effective interaction \( V(\vec{r}_{1}-\vec{r}_{2}|) \),
which diverges only logarithmically as \( r\rightarrow 0 \), will
be significantly smaller than those of the bare Coulomb interaction,
the reduction of the gap due to Landau level mixing in these samples
should, in fact, be significantly smaller than this figure of 10\%
and is probably negligible. Clearly, a new study along the lines of
\cite{Rez_LL_mixing} (which actually looked at the harder problem
of Landau-level mixing at \( \nu =5/2 \) for systems with a partially
filled second Landau level), taking account of the finite-width of
the subband wf, would make for significantly more accurate estimates
of Landau level mixing effects.

We should also consider the role of excitations involving spin reversals.
The gap at \( \nu =1/3 \), corresponding to the creation of a quasihole
with no spin-reversal and a quasiparticle with one spin-reversal,
was estimated in \cite{morf_ha87,Rez_spin_reversals_91} using extensive
Monte Carlo simulations of trial wf's. Gap estimates for various combinations
of quasihole and quasiparticles combined with spin-reversals based
on exact diagonalizations of small systems (up to six particles) were
reported in \cite{chakraborty_86}. For the case of the pure Coulomb
interaction and ignoring the Zeeman energy the gap to create spin-reversed
excitations was around 60\% smaller than the spin-polarized gaps for
systems at \( \nu =1/3. \) When the Zeeman energy (in GaAs) is taken
into account this suggests that, for a pure Coulomb interaction, the
spin-reversed excitation would have a lower energy for systems at
\( \nu =1/3 \) if the magnetic field were smaller than \( \sim 7 \)T.
This is above the fields at which the \( 4/3 \) and \( 5/3 \) states
were observed in \cite{willett88} (see Table \ref{Willett_data}),
and may account for the larger discrepancy seen at these filling fractions
than at \( \nu =1/3 \) or \( 2/3 \).

The estimate of 7T, as the field below which the spin-reversed excitation
drops below the spin polarized excitation, is well below the fields
in Table \ref{Du_A&B} making it unlikely that spin-reversed excitations
are involved at these filling fractions. Although the explicit estimate
of the spin-reversed excitation was made for a system at \( \nu =1/3 \)
it is unlikely that the discrepancy at other filling fractions will
be larger. This is because the difference between the spin-polarized
and spin-reversed quasiparticle energies should be largest at \( \nu =1/3 \),
where it is possible to construct a spin-reversed quasiparticle state
which is a zero-energy eigenstate of the hard-core potential (with
only the pseudopotentials \( V_{0} \) and \( V_{1} \) non-zero).
For the case of the Coulomb interaction, its energy is controlled
by the size of the pseudopotential \( V_{2} \), while the energy
of the spin-polarized quasiparticle is determined by the larger \( V_{1} \).
However, as \( V_{2} \) is reduced less by finite width effects than
\( V_{1} \) (see Figure \ref{Delta V(L)}), the spin-polarized quasiparticle
will be stabilized with respect to the spin-reversed excitation by
finite-width effects \cite{morf_ha87}.

The results for the activation gaps at \( \nu =1/3 \) and \( 2/5 \)
in layers of tetracene reported in \cite{tetracene_gaps} can also
be compared with our numerical results. By simultaneously varying
the gate voltage and magnetic field the gaps could be tracked as a
function of the ratio \( w/\ell _{0} \) for a range of fields \( 3<B[T]<9 \).
One intriguing feature of these organic layers is that the ratio of
the Coulomb interaction to Landau level spacing, \( (e^{2}/\varepsilon \ell _{0})/\hbar \omega _{c} \),
is approximately 30-40 times larger than in the GaAs samples for systems
at the same magnetic fields. 

We have computed the width, \( w \), of the subband wavefunctions
in the tetracene samples of \cite{tetracene_gaps} excluding the effects
of image charges using the zG trial wavefunction and found that \( w \)
varies between \( 17\AA  \) at a density \( n_{s}=0.1\times 10^{11} \)
cm\( ^{-2} \) and \( 7\AA  \) for \( n_{s}=5\times 10^{11} \)cm\( ^{-2} \).
The effects of the image potential will be to reduce the width still
further. At all the densities, at which the gaps were measured, \( w/\ell _{0}<0.1 \)
and so the effects of the finite width of the wavefunction on the
gaps in these samples are small (see Fig \ref{gaps(width)}) and significantly
smaller than were reported in Reference \cite{tetracene_gaps,schoen_epsilon}.
However, our calculations summarized in Fig \ref{gaps(width)} are
a more accurate reflection of finite width effects than the old formula
of \cite{zhang_das86} used in \cite{tetracene_gaps}. We also note
that the widths we obtain are about half the order of magnitude (\( w\sim 35\AA  \))
quoted in \cite{tetracene_gaps}. 

\begin{figure}
{\centering \resizebox*{8cm}{!}{\includegraphics{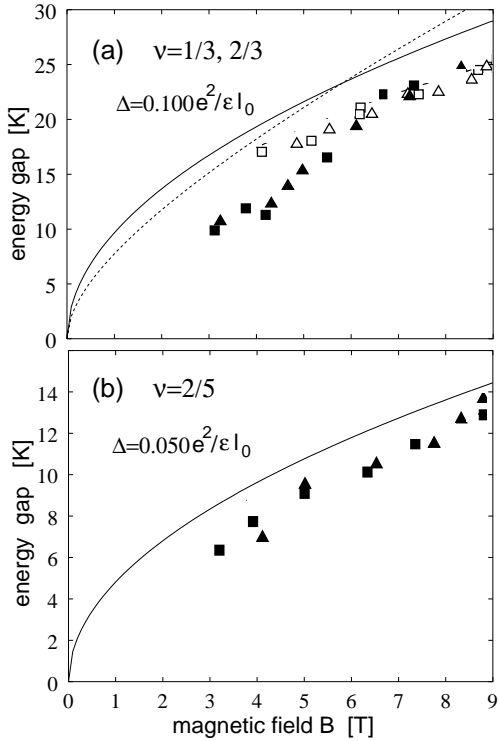}} \vskip -8 truemm\par}

\caption{\label{tetracene_fig}Gap energies at \protect\( \nu =1/3\protect \)
(upper panel, unfilled symbols) and \protect\( 2/3\protect \) (upper
panel, filled symbols) and \protect\( 2/5\protect \) (lower panel)
in tetracene samples from \cite{tetracene_gaps} compared with numerical
estimates. Triangles and squares denote experimental results for different
samples. The solid line in each panel shows the gap given by the quoted
formula which is valid in the zero thickness limit (as discussed in
the text the effects of the non-zero width are negligible in these
samples). The effective dielectric constant is taken as the average
\protect\( \epsilon =6.65\protect \) of reported values for tetracene
\protect\( \epsilon =3.5\protect \) and the value \protect\( \epsilon =9.8\protect \)
for the alumina insulating layer. The dashed line in the upper panel
is an estimate of the energy gap for
spin reversed excitations (see text).}
\end{figure}

In order to compare the results of our calculations with the results
of the experiments on the tetracene MIS structures, we need to take
account of the large difference between the dielectric constant of
the alumina insulating layer (\( \varepsilon \sim 9.8) \) and the
value for tetracene (\( \varepsilon \sim 3.5 \)). Nearly all the
charge density is within \( 3w \) of the interface. This is significantly
less than the average interparticle spacing or ion disk radius (\( a=\sqrt{{2/\nu }}\ell _{0} \)),
which for these samples varies from 150\( \AA  \) to 360\( \AA  \)
depending on density. We have used the simplest approximation 
which  treats
the 2D electron gas as if it were trapped at the interface of the
alumina and the semiconductor. The effective dielectric
constant is then just the average for the two materials. (Corrections to
this, taking account of the actual displacement of the charge away
from the interface, would involve image charge effects and give rise to
a change in the functional form of the effective interaction between
particles \cite{jackson_99}.) In Figure \ref{tetracene_fig}, we compare
the results we obtain with the measured values reported in \cite{tetracene_gaps}.
We show calculated gaps as a function of magnetic field ignoring the
finite width of the charge distribution. 

The difference between the computed and measured values of the gaps
in Figure \ref{tetracene_fig} at \( \nu =1/3 \) and \( 2/5 \) are
remarkably small. Although there is some uncertainty associated with
the computed gaps arising from the simple treatment of the large difference
in dielectric constant of tetracene and alumina, 
there is surprisingly little evidence of large Landau level mixing (LLM) or
disorder-related effects at
these filling fractions. The ratio of the Coulomb energy scale
$e^2/\varepsilon \ell_0$
to the cyclotron energy in tetracene is $\sim 93/\sqrt{B}$ or $\sim 42$ 
at $B=5$T and LLM  should be significant and might even be expected to be
dramatic. When the ratio of these two
energy scales is this large
a perturbative treatment of LLM effects may not even be possible. 
(We have assumed the same effective
dielectric constant as used in Fig \ref{tetracene_fig}). 
Even though the mobilities are not as high
in the tetracene MIS structures as in the GaAs-GaAlAs heterostructures
\cite{du93,willett88}, the agreement between calculated and measured
gaps suggest that there are not any strong effects of disorder scattering
either.  

For the systems at \( \nu =2/3 \) (filled symbols in upper panel
of Fig \ref{tetracene_fig}), it was suggested in \cite{tetracene_gaps}
that the change in slope at around 6.5T was related to a transition
from a polarized state at high fields to a state which was not fully
polarized at lower fields. This seems unlikely. At a transition with
a discontinuity in polarization (first order transition), there would
normally be a discontinuity in the gap rather than a discontinuity
in its gradient with magnetic field, see for example \cite{Eisenstein_8/5}.
There has been one report of a transition from a polarized to partially
polarized state at \( \nu =2/3 \) in GaAs heterostructures without
any discontinuity of the gap \cite{Freytag}. However the corresponding
transition would be expected to occur in the tetracene samples at
around 1.7T well below the range of fields of Fig \ref{tetracene_fig}.
Even if there were no (or only a small) discontinuity in the gap,
the change in slope would normally be in the opposite sense to the
one reported (see Figure \ref{tetracene_fig}). The lowest-lying excitations
from a partially polarized (or unpolarized) state would be expected
to involve spin reversals which increased the total aligned spin (rather
than reduced it) and thereby gained a reduction in Zeeman energy.
On the other hand, excitations from the fully polarized state, either
decrease the total spin or leave it constant. As a result there
would either be a contribution to the energy of the excitation from
the Zeeman energy, which was positive and increasing as a function
of field, or no contribution. In either case, the gap would be expected
to grow faster with field in the high field (fully polarized) state
than in the low field state but not not more slowly as reported in
\cite{tetracene_gaps}. This is what was observed for the transition
seen at \( \nu =8/5 \) \cite{Eisenstein_8/5}. 

An alternative explanation of the results at \( \nu =2/3 \), assumes
a fully polarized ground state and identifies the change in slope
at \( B\approx 6.5 \)T with a change in the nature of the lowest
lying excitations. For \( B\gtrsim 6.5 \)T, the lowest energy excitations
would be within the fully spin-polarized sector, while for \( B\lesssim 6.5 \)T
they would involve a spin-reversal. We can make a rough estimate of
the energy of the spin-reversed excitation gap ignoring Landau level
mixing (LLM) as follows. Previous estimates of the Coulomb energy
of a spin-reversed quasiparticle energy put it at around 55\% of the
energy of the spin-polarized quasiparticle \cite{morf_ha87}. The
Coulomb energy of a spin-reversed quasihole is unlikely to be much
lower than that of the spin-polarized hole, while the additional Zeeman
energy will make this excitation unfavorable. We therefore take for
the value of the quasihole energy that of the spin-polarized hole.
The results for this `spin-reversed' energy gap are shown as a dashed
line in Figure (\ref{tetracene_fig}) and are seen to be quite close
to the observed data points, although the difference between our results
for the `polarized' gap and the spin reversed' gap is small. Our estimate
of the spin-reversed excitation applies both at \( \nu =1/3 \) and
\( \nu =2/3 \) as we have neglected LLM, which allows electrons to
make virtual transitions to other Landau levels and thereby screen
the interaction in the lowest Landau level. These effects would be
larger for systems at \( \nu =2/3 \) than at \( \nu =1/3 \) in the
same magnetic field and could explain why the the spin-reversed excitation
lies below the fully polarized excitation up to higher magnetic fields
at \( \nu =2/3 \) than at \( \nu =1/3 \). A tilted field experiment
\cite{Eisenstein_8/5} would be one method to determine whether our
identification of the change of slope in the gap with field with a
change in polarization of the lowest-lying excitation is correct.

The apparent absence of a significant reduction of the gap in the
tetracene MIS structures (\( \mu <2.5\times 10^{5} \) cm\( ^{2} \)/Vs)
on account of disorder, given its importance in the ultra-high mobility
(\( 12.8\times 10^{6} \) cm\( ^{2} \)/Vs) GaAs heterostructures,
is puzzling. It suggests that the activated gap measured in transport
measurements is affected by disorder in different ways in heterostructures
and MIS structures. In the heterostructures, the disorder scattering
is that of the ionized silicon donors which were in a layer about
800\( \AA  \) from the electrons \cite{du93}. In the MIS structures,
on the other hand, the doping is controlled by a capacitance (\( \sim 130 \)
nFcm\( ^{-2} \) \cite{tetracene_gaps}) with the backgate of order
microns from the carriers. Here the disorder scattering is likely
to be that of neutral defects. It is possible that, in the heterostructures,
the activation studies do not measure directly the energy to create
a quasiparticle quasihole pair from the ground state, but rather the
energy to excite quasiparticle (or quasiholes) out of bound states
in the potential of the (charged) impurity distribution.

\section{Conclusions}

We have used diagonalizations of the Hamiltonians for finite size
systems on a sphere to obtain estimates of the gaps at filling fractions
in the Jain sequence \( \nu =1/3,\, 2/5,\, 3/7 \) and 4/9 and at
\( \nu =5/2 \). We have emphasized how taking account properly of
the systematic contributions to the excitation energy from the charge
redistribution on the sphere in excited states is essential if one
is to obtain accurate estimates of the gaps in the thermodynamic limit.
Our results for the gaps are smaller than earlier estimates based
on finite-size studies (for \( \nu =2/5 \) and 3/7 \cite{nda_morf89})
and those based on the study of trial wavefunctions (for \( \nu =2/5,\, 3/7 \)
and 4/9 \cite{jain_kamilla97}). This difference is important, as
estimates of the gap as a function of \( \nu  \) provide the most
direct numerical estimates the effective mass of CF's \cite{HLR,Bert_in_Perspectives}.
Our new results are consistent with the CF picture provided the logarithmic
corrections to the effective mass are taken into account and are not
well described by assuming a filling factor independent effective
mass (see Fig \ref{fig:eff_mass(p)}).

We have shown that Gaussian and the z\( \times  \)Gaussian (zG) variational
functions accurately describe subband wavefunctions and yield subband
energies and lowest Landau level pseudopotentials, which are essentially
indistinguishable from those obtained by solving for the subband wavefunctions
exactly by direct numerical integration. These trial wavefunctions
offer a significant improvement over the standard Fang-Howard (FH)
form, which overestimates the standard deviation of the charge distribution,
\( w \), by as much as 50\% depending on electron density, \( n_{s} \),
(see Fig \ref{Trial wavefunctions compared}). The lowest Landau level
pseudopotentials, which are the starting point for the study of the
fractional quantum Hall gaps, turn out to be accurately determined
using any of the three trial forms (Gaussian, zG or FH) once \( w \)
is known (see Fig \ref{Delta V(L)}). This offers a huge simplification
over the previous \emph{ab initio} approaches which used numerical
integration to find the subband wavefunction and tables of pseudopotential
parameters \cite{ortalano97}. We have also computed the variation
of the gaps at fractionally quantized Hall states as a function of
width of the subband charge distribution. The results are parametrized
in Eq \ref{eq:interpol} and Table \ref{table:interpol}. 

We have compared our computed gaps with measured activated gaps. We
have found that, even after taking account of disorder broadening
of states, the measured activation gaps in GaAs heterostructures are
only around 60\% of the computed gaps for the filling fractions \( \nu =1/3,\, 2/5,\, 3/7 \)
and 4/9 (see Table \ref{Du_A&B}). This is to be contrasted with the
activated gaps at \( \nu =1/3 \) and 2/5 reported in tetracene MIS
structures, which turn out to be reasonable agreement with computed
gaps (see Fig \ref{tetracene_fig}). We have suggested that the relationship
between the computed gap and the measured activated gap may be different
depending on the type of disorder in the samples. In the GaAs heterostructures
the charged donor ions, which are the main scattering centers, are
only around 800\( \AA  \) from the quantum Hall layer, and this could
lead to local variations in the energy required to excite quasiparticle-quasihole
pairs with the lowest excitation energies controlling the activated
transport. On the other hand, the backgate in the MIS structures is
of order microns from the quantum Hall layer, and the main scattering
centers are likely to be neutral. These are less likely to affect
the energy to excite quasiparticle-quasihole pairs and the gap controlling
activated transport should then be close to the true thermodynamic
gap as we have found. For filling fractions \( \nu =2/3 \) and \( 5/2 \),
the reported activated transport gaps in GaAs heterostructures are
only around 10\% and 5\% respectively of the values we compute, although
we have not attempted to account for disorder broadening in these
cases. However, the experimental evidence suggests a disorder broadening
which is comparable at \( \nu =5/2 \) with the computed gap \cite{pan_exact_quantization}
so the large discrepancy is to be expected.

\begin{acknowledgments}
We would like to thank B.I. Halperin for helpful discussions. Two
of us (RHM and SDS) would like to thank the ITP in Santa Barbara for
its hospitality in 1998, which led directly to this project, and the
Aspen Center for Physics for its hospitality in 2000 where this project
made important progress.
\end{acknowledgments}

\section*{Appendix: ZdS Interaction}

\begin{figure}
{\centering \resizebox*{7cm}{!}{\includegraphics{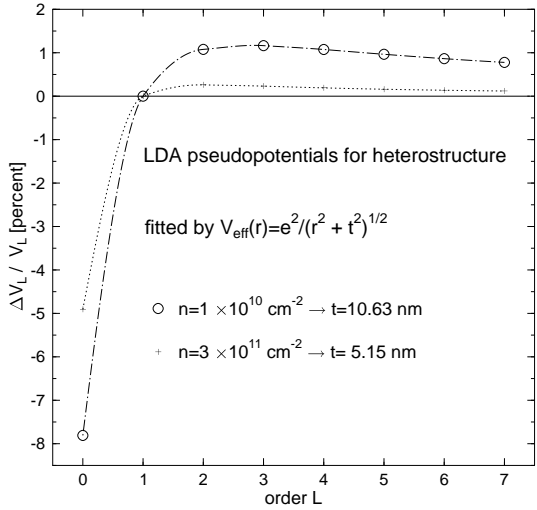}} \par}

\caption{\label{fig:zds}Errors in the Haldane pseudopotentials \protect\( V_{L}\protect \)
computed using the Zhang-DasSarma model interaction (\ref{Zhang_das_Sarma_Interaction}).
The comparison is with the values reported in \cite{ortalano97}.
The variational parameters are determined such that \protect\( V_{1}\protect \)
is correctly reproduced. Note in particular the large errors for \protect\( V_{0}\protect \)
and the slow decay of the error with increasing \protect\( L\protect \)
for the low-density sample.}
\end{figure}

One previous attempt to model finite width effects used the `model
interaction' \cite{zhang_das86} \begin{equation}
\label{Zhang_das_Sarma_Interaction}
V_{ZdS}(r)=\frac{e^{2}}{\varepsilon }\frac{1}{\sqrt{r^{2}+t^{2}}},
\end{equation}
which introduces a width parameter \( t \). We have found that this
model interaction cannot reproduce accurately the variation with \( L \)
of the Haldane pseudopotential parameters for a sample with finite
width with the errors significantly increasing as the width increases,
Figure \ref{fig:zds}.

\begin{figure}
{\centering \resizebox*{7cm}{!}{\includegraphics{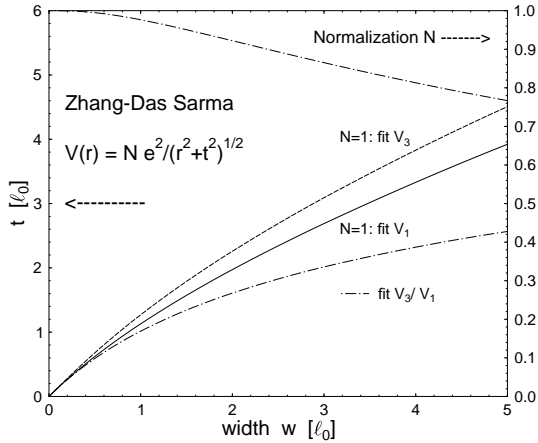}} \par}

\caption{\label{ZdS_renormalized}The width parameter \protect\( t\protect \)
(lower three curves) in the ZDS model interaction in(\ref{ZdS_with_normalization})
and the normalization parameter \protect\( N\protect \) (top curve),
as a function of the Gaussian width \protect\( w\protect \) of the
density distribution, \protect\( \zeta (z)^{2}\protect \). For the
two curves labelled \protect\( N=1,\protect \) the normalization
is held equal to one and the parameter \protect\( t\protect \) is
chosen so that the Haldane pseudopotential \protect\( V_{1}\protect \)(lower
curve) and \protect\( V_{3}\protect \) (upper curve) for the interaction
in \ref{ZdS_with_normalization} are equal to the values obtained
from (\ref{2D from 3D interaction}) using the Gaussian variational
wf's. The dashed-dotted curves show the values of \protect\( t\protect \)
and \protect\( N\protect \) required to reproduce both \protect\( V_{1}\protect \)
and \protect\( V_{3}/V_{1}\protect \) correctly. }
\end{figure}

The reason for this is probably the unphysical nature of this interaction
as a model for electrons in a heterostructure or quantum well interacting
via the Coulomb interaction. Taking the Fourier transform of (\ref{2D from 3D interaction})
and using the convolution theorem, one can show that it is not possible
to construct a density distribution, \( |\zeta (z)|^{2} \), for which
the effective interaction (see\ref{2D from 3D interaction}) is \( V_{ZdS}(r) \).
This is essentially because \( V_{ZdS}(r) \) is the Coulomb interaction
of two (distinguishable) particles confined to separate planes a distance
\( t \) apart and, as such, misses the \( \ln r \) found for small
\( r \) and large widths for all realistic density distributions,
\( |\zeta (z)|^{2} \). However, many of the results obtained on the
basis of the effective interaction are still valid if interpreted
carefully.

For this purpose, we incorporate an overall scaling factor of the
interaction \( N \),\begin{equation}
\label{ZdS_with_normalization}
v_{ZdS}(r)=N\frac{e^{2}}{\varepsilon }\frac{1}{\sqrt{r^{2}+t^{2}}},
\end{equation}
with \( N=1 \) giving the original interaction (\ref{Zhang_das_Sarma_Interaction}).
The gap energies and relative stability of fractional quantum Hall
states in the principal Jain sequence are determined principally by
the first two Haldane pseudopotentials for odd angular momentum \( V_{1} \)
and \( V_{3} \). In figure \ref{ZdS_renormalized} we show the values
of \( t \) required in \ref{ZdS_with_normalization} to match the
values of \( V_{1} \) and of \( V_{3} \) to those obtained using
the variational Gaussian wf as a function of the width parameter assuming
\( N=1. \) It is clear that it is not possible to find a value of
\( t \) which gives both \( V_{1} \) and \( V_{3} \) correctly.
If we allow \( N \) and \( t \) to vary then both \( V_{1} \) and
\( V_{3} \) can be correctly reproduced by the effective interaction
in (\ref{ZdS_with_normalization}). The results are also shown in
Fig \ref{ZdS_renormalized}. Changing \( N \) means that the asymptotic
behavior of the pseudopotentials at large angular momentum is not
reproduced correctly. However, as the gaps and stability of the incompressible
states in the Jain sequence are determined principally by the pseudopotentials
\( V_{1} \)and \( V_{3} \) this should not be a problem. If phase
transitions between spin-singlet and polarized states (e.g. at \( \nu =2/5) \)
are of interest, it is obviously possible to correctly represent the
in this case most important pseudopotentials \( V_{1} \) and \( V_{2} \)
for angular momenta \( L=1 \) and \( 2 \), by appropriate choice
of \( N \) and \( t. \)


\begin{thebibliography}{10}
\bibitem{schoen_epsilon}The results for the theoretical values of the energy gaps at \( \nu =1/3 \)
and 2/5 reported in Reference \cite{tetracene_gaps} are apparently
based on different values of the dielectric constant, \( \epsilon =5.86 \)
at \( \nu =1/3 \) and \( \epsilon =6.6 \) at \( \nu =2/5 \). This
is not discussed in Reference \cite{tetracene_gaps}.
\end{thebibliography}
\end{document}